%% file: oraclepaper.tex
\documentclass[sigconf]{acmart}
\pdfoutput=1

\AtBeginDocument{%
  }

\copyrightyear{2024}
\acmYear{2024}
\setcopyright{licensedcagovmixed}
\acmConference[ICSE '24]{2024 IEEE/ACM 46th International Conference on Software Engineering}{April 14--20, 2024}{Lisbon, Portugal}
\acmBooktitle{2024 IEEE/ACM 46th International Conference on Software Engineering (ICSE '24), April 14--20, 2024, Lisbon, Portugal}

\usepackage{listings}
\usepackage{multicol}
\usepackage{amsmath}

\usepackage{amssymb}
\usepackage{listings}
\usepackage{graphicx}
\usepackage{makecell}
\usepackage{algorithm}
\usepackage[noend]{algpseudocode}
\usepackage{adjustbox}
\usepackage{tabularx}
\usepackage{adjustbox}
\usepackage{booktabs, multirow} 
\usepackage{soul}
\usepackage{changepage,threeparttable} 
\usepackage[justification=centering]{caption}
\usepackage{multirow}

\usepackage{enumitem}

\newcommand\CONDITION[2]%
  {\begin{tabular}[t]{@{}l@{}l@{}}
     #1&#2
   \end{tabular}%
  }
\algdef{SE}[WHILE]{While}{EndWhile}[1]%
  {\algorithmicwhile\ \CONDITION{#1}{\ \algorithmicdo}}%
  {\algorithmicend\ \algorithmicwhile}
\algdef{SE}[FOR]{For}{EndFor}[1]%
  {\algorithmicfor\ \CONDITION{#1}{\ \algorithmicdo}}%
  {\algorithmicend\ \algorithmicfor}
\algdef{S}[FOR]{ForAll}[1]%
  {\algorithmicforall\ \CONDITION{#1}{\ \algorithmicdo}}
\algdef{SE}[REPEAT]{Repeat}{Until}{\algorithmicrepeat}[1]%
  {\algorithmicuntil\ \CONDITION{#1}{}}
\algdef{SE}[IF]{If}{}[1]%
  {\algorithmicif\ \CONDITION{#1}{\ \algorithmicthen}}%

\algdef{C}[IF]{IF}{ElsIf}[1]%
  {\algorithmicelse\ \algorithmicif\ \CONDITION{#1}{\ \algorithmicthen}}

\newcommand*{\name}{{OVer}}

\newcommand*{\rev}[1]{{#1}}

\newcommand*{\revision}[1]{{#1}}
\newcommand*{\dsl}[1]{\mathbb{#1}}
\newcommand*{\ssol}[1]{\mathcal{#1}}
\newcommand*{\summary}[1]{\mathbb{#1}}

\newcommand*{\replace}[3]{{#1}{[#2 {/}}{#3]}}

\algnewcommand\algorithmicforeach{\textbf{for each}}
\algdef{S}[FOR]{ForEach}[1]{\algorithmicforeach\ #1\ \algorithmicdo}

\usepackage{xcolor}
\definecolor{LightGray}{gray}{0.9}
\definecolor{codegreen}{rgb}{0,0.6,0}
\definecolor{codegray}{rgb}{0.5,0.5,0.5}
\definecolor{codepurple}{rgb}{0.58,0,0.82}
\definecolor{backcolour}{rgb}{1,1,1}

\lstdefinestyle{mystyle}{
    backgroundcolor=\color{backcolour},   
    commentstyle=\color{codegreen},
    keywordstyle=\color{magenta},
    numberstyle=\tiny\color{codegray},
    stringstyle=\color{codepurple},
    basicstyle=\ttfamily\footnotesize,
    breakatwhitespace=false,         
    breaklines=false,                 
    captionpos=b,                    
    keepspaces=true,                 
    numbers=left,                    
    numbersep=5pt,                  
    showspaces=false,                
    showstringspaces=false,
    showtabs=false,                  
    tabsize=2
}

\input{solidity-highlight}

\begin{document}

\title{Safeguarding DeFi Smart Contracts against Oracle Deviations}

\author{Xun Deng}
\email{xun.deng@mail.utoronto.ca}
\orcid{0009-0003-9364-1412}
\affiliation{
  \institution{University of Toronto}
  \city{Toronto}
  \country{Canada}
}
\author{Sidi Mohamed Beillahi}
\email{sm.beillahi@utoronto.ca}
\orcid{0000-0001-6526-9295}
\affiliation{
  \institution{University of Toronto}
  \city{Toronto}
  \country{Canada}
}

\author{Cyrus Minwalla}
\email{cminwalla@bank-banque-canada.ca}
\orcid{0000-0002-9569-664X}
\affiliation{
  \institution{Bank of Canada}
  \city{Ottawa}
  \country{Canada}
}

\author{Han Du}
\email{HDu@bank-banque-canada.ca}
\orcid{0009-0005-0256-180X}
\affiliation{
  \institution{Bank of Canada}
  \city{Ottawa}
  \country{Canada}
}

\author{Andreas Veneris}
\email{veneris@eecg.toronto.edu}
\orcid{0000-0002-6309-8821}
\affiliation{
  \institution{University of Toronto}
  \city{Toronto}
  \country{Canada}
}

\author{Fan Long}
\email{fanl@cs.toronto.edu}
\orcid{0000-0001-7973-1188}
\affiliation{
  \institution{University of Toronto}
  \city{Toronto}
  \country{Canada}
}

\input{abstract}



\maketitle

\input{introduction}
\input{background}

\input{overview}
\input{methodology}
\input{experiment}
\input{relatedwork}
\input{threat}
\input{conclusion}

\bibliographystyle{ACM-Reference-Format}
\bibliography{main}

\end{document}

%% file: solidity-highlight.tex
\usepackage{listings, xcolor}

\definecolor{verylightgray}{rgb}{.97,.97,.97}

\lstdefinelanguage{Summary}{
  sensitive = true,
  backgroundcolor=\color{white}
  keywords={},
  otherkeywords={
    >, <, ==, mul, add, sub, div, +, -, /, *
  },
  keywords = [2]{SUM, from, to, sum, index},
  keywordstyle=\color{blue},
  keywordstyle=[2]\color{magenta},
  numbers=left,
  numberstyle=\scriptsize,
  stepnumber=1,
  numbersep=8pt,
  showstringspaces=false,
  breaklines=true,
  frame=none,
  comment=[l]{//},
  morecomment=[s]{/*}{*/},
  commentstyle=\color{purple}\ttfamily,
  stringstyle=\color{red}\ttfamily,
  morestring=[b]',
  morestring=[b]"
  }

\lstdefinelanguage{Solidity}{
	keywords=[1]{anonymous, assembly, assert, balance, break, call, callcode, case, catch, class, constant, continue, constructor, contract, debugger, default, delegatecall, delete, do, else, emit, event, experimental, export, external, false, finally, for, function, gas, if, implements, import, in, indexed, instanceof, interface, internal, is, length, library, log0, log1, log2, log3, log4, memory, modifier, new, payable, pragma, private, protected, public, pure, push, require, return, returns, revert, selfdestruct, send, solidity, storage, struct, suicide, super, switch, then, this, throw, transfer, true, try, typeof, using, value, view, while, with, addmod, ecrecover, keccak256, mulmod, ripemd160, sha256, sha3}, 
	keywordstyle=[1]\color{blue}\bfseries,
	keywords=[2]{address, bool, byte, bytes, bytes1, bytes2, bytes3, bytes4, bytes5, bytes6, bytes7, bytes8, bytes9, bytes10, bytes11, bytes12, bytes13, bytes14, bytes15, bytes16, bytes17, bytes18, bytes19, bytes20, bytes21, bytes22, bytes23, bytes24, bytes25, bytes26, bytes27, bytes28, bytes29, bytes30, bytes31, bytes32, enum, int, int8, int16, int24, int32, int40, int48, int56, int64, int72, int80, int88, int96, int104, int112, int120, int128, int136, int144, int152, int160, int168, int176, int184, int192, int200, int208, int216, int224, int232, int240, int248, int256, mapping, string, uint, uint8, uint16, uint24, uint32, uint40, uint48, uint56, uint64, uint72, uint80, uint88, uint96, uint104, uint112, uint120, uint128, uint136, uint144, uint152, uint160, uint168, uint176, uint184, uint192, uint200, uint208, uint216, uint224, uint232, uint240, uint248, uint256, var, void, ether, finney, szabo, wei, days, hours, minutes, seconds, weeks, years},	
	keywordstyle=[2]\color{teal}\bfseries,
	keywords=[3]{block, blockhash, coinbase, difficulty, gaslimit, number, timestamp, msg, data, gas, sender, sig, value, now, tx, gasprice, origin},	
    keywordstyle=[3]\color{violet}\bfseries,
	identifierstyle=\color{black},
	sensitive=true,
	comment=[l]{//},
	morecomment=[s]{/*}{*/},
	commentstyle=\color{gray}\ttfamily,
	stringstyle=\color{red}\ttfamily,
	morestring=[b]',
	morestring=[b]"
}

%% file: abstract.tex
\begin{abstract}
    This paper presents {\name}, a framework designed to automatically analyze the behavior of decentralized finance (DeFi) protocols when subjected to a "skewed" oracle input. 
    {\name} firstly performs symbolic analysis on the given contract and constructs a model of constraints. Then, the framework leverages an SMT solver to identify parameters that allow its secure operation. 
    Furthermore, guard statements may be generated for smart contracts that may use the oracle values, thus effectively preventing oracle manipulation attacks. 
    Empirical results show that {\name} can successfully analyze all 10 benchmarks collected, which encompass a diverse range of DeFi protocols. 
    Additionally, this paper also illustrates that current parameters utilized in the majority of benchmarks are inadequate to ensure safety when confronted with significant oracle deviations.
\end{abstract}

%% file: introduction.tex
\section{Introduction}

Blockchain offers decentralized, programmable, and robust ledgers on a global
scale. Smart contracts, which are programs deployed on blockchains, encode
transaction rules to govern these blockchain ledgers. This technology has been
adopted across a wide range of sectors, including financial services, supply
chain management, and entertainment. A notable application of smart contracts
is in the management of digital assets to create decentralized financial
services (DeFi). As of April 1st, 2023, the Total Value Locked (TVL) in 1,417
DeFi contracts had reached \$50.15 billion~\cite{defillamaref}.

As the assets managed by smart contracts continue to grow, ensuring their
correctness has become a critical issue. In response, researchers have
developed numerous analysis and verification tools to detect errors in contract
implementation. However, beyond the typical software challenges posed by
implementation errors, the correctness of many DeFi smart contracts often
depends on \emph{oracle values}~\cite{adler2018astraea}. These are external values that capture vital
environmental conditions under which the contracts operate. For instance, a
collateralized DeFi lending contract requires updated trading prices of various
digital assets to ensure that the value of the collateral asset always exceeds
the value of the borrowed asset for each user.

Smart contracts periodically receive updates to their oracle values from other
contracts or external databases and APIs. Deviations in these oracle values
from their true values can lead to deviations in the intended operations of the
contracts~\cite{cai2022truthful,adler2018astraea}. In the real world, such
deviations are common, often stemming from inaccuracies in the value source or
delays in transmission. DeFi protocols traditionally use a variety of empirical
strategies to mitigate the risks associated with oracle deviations and
potential corruptions. For instance, a leveraged DeFi protocol might set a
safety margin for user positions, liquidating a position if its asset price
dips below a specific threshold. Alternatively, a protocol might aggregate
multiple oracle inputs from varied sources, calculating a median or average for
computational purposes. However, these mechanisms and their parameters are
often ad-hoc and arbitrary. The adequacy and efficacy of these control
mechanisms in real-world scenarios remain uncertain.

This paper presents {\name}, the first sound and
automated tool for analyzing oracle deviation and verifying its impact in DeFi
smart contracts. Given the source code of a smart contract protocol and a deviation range
of specific oracle values in the contracts, {\name}
automatically analyzes the source code to extract a summary of the protocol.
For a safety constraint of the protocol, {\name} then uses the extracted
summary to determine how to appropriately set key control parameters in the contract.
This ensures that the resulting contract continues to satisfy the desired
constraint, even in the face of oracle deviations.

One of the key challenges {\name} faces is the sophisticated contract logic of
DeFi protocols. DeFi contracts often contain multiple loops that iterate over
map-like data structures. Each iteration typically contains up to a hundred
lines of code to handle the protocol logic for one kind of asset or for one
user account. Such code patterns are typically intractable for standard program
analysis techniques, which often would have to make undesirable
over-approximation or to bound the number of loop iterations, leading to
inaccurate or unsound analysis results.

{\name} tackles this challenge with its innovative loop summary algorithm.
Since the essence of loops computations in DeFi protocols consists of
accumulators applied to maps data structures, {\name} operates with a
predefined sum operator template for loops. {\name} extracts the summary
formula of each iteration and then uses a template-based approach to convert
the extracted expressions into an instantiation of the sum operator template to
represent the summary of the entire loop. Distinct from previous loop summary
algorithms that struggle with complex if-else branching or multifaceted folding
operations with interdependencies~\cite{loopSummary}, the OVer algorithm
adeptly manages these prevalent complexities in popular DeFi contracts.

We have evaluated {\name} on a set of
nine popular DeFi protocols and one fictional protocol in our experiments. 
{\name} successfully analyzes all the protocols, each taking less than nine seconds.
In comparison, a prior state-of-the-art loop summary algorithm can only handle 
0 out of 7 benchmarks that have loops.

With {\name}, we study the history oracle deviation in real-world blockchains.
We investigate how oracle deviation would affect the behavior of popular DeFi
contracts and whether the existing ad-hoc mechanisms are sufficient to
neutralize the oracle deviations. Our results show that for six out of the
seven benchmark protocols, the control mechanism was insufficient to handle the
oracle deviation for at least a certain period of time, leading to temporary
exploitable vulnerabilities. Our results also surprisingly show that existing
ad-hoc mechanisms often exacerbate the security issue caused by oracle
deviation. For example, to protect against potentially malicious oracle value
providers, several DeFi protocols introduce delays when using oracle value
inputs in their calculation (\emph{e.g.}, using the reported asset price one
hour ago as the current oracle price). When the digital asset price fluctuates,
such mechanisms fail to reflect the current market and artificially inject
deviations, which may make the resulting protocols more vulnerable.

In summary, this paper makes the following contributions.
\begin{itemize}[noitemsep,topsep=0pt]
\item \emph{{\name}:} This paper presents {\name}, the first sound analysis and verification
tool for analyzing oracle deviation in DeFi protocols.

\item \emph{Loop Summary Algorithm:} This paper proposes a novel loop summary
algorithm to enable the analysis of the sophisticated loops in DeFi smart
contract source code. 

\item \emph{Results:} This paper presents a systematic evaluation of {\name}.
It also presents the first study of oracle deviation on popular DeFi protocols.
Our results show that the existing ad-hoc control mechanisms are often
insufficient or even detrimental to protect the DeFi protocols against the
oracle deviation in the real-world.
\end{itemize}

The remaining of the paper is organized as follows.
Sections~\ref{sec:background}-\ref{sec:example} introduce technical background and a motivating example.
Section~\ref{sec:design} presents the design of {\name}. 
In Section~\ref{sec:results}, we study past oracle deviations and evaluate {\name}. 
We discuss related work and threats to the validity in Sections~\ref{sec:related_work}-\ref{sec:threat} and conclude in Section~\ref{sec:conclusion}.

%% file: background.tex
\section{Background} \label{sec:background}

\emph{Blockchain and smart contracts.} 
Blockchains, operating as decentralized distributed systems, offer a formidable
architecture for resilient, programmable ledgers. Numerous blockchain
infrastructures, with Ethereum as a prime example, provide support for smart
contracts. These are coded agreements residing on the blockchain, established
to administer transaction rules integral to ledger operations. Commonly
scripted in sophisticated languages such as \emph{Solidity}~\cite{solc}, these smart contracts are
later compiled into a lower-level machine language like Ethereum Virtual
Machine (EVM) bytecode~\cite{buterin2014ethereum}. For consistent enforcement of these transaction rules,
all participating nodes within the blockchain network execute the bytecode of a
contract in a consensus-oriented fashion.

\emph{Decentralized finance protocols.} 
A substantial application of block- chain technology is visible in the form of DeFi protocols. DeFi protocols
deploy smart contracts to manage digital assets, enabling an array of financial
services encompassing trading, lending, and investment, all within a
decentralized context. Predominantly, DeFi applications consist of automatic
market makers (AMMs) and lending protocols, with AMMs being a frequent
component of decentralized exchanges (DEXes).

Contrasting traditional exchanges that utilize order books for trading
operations, AMMs implement a mathematical model that is contingent on the
asset's volume in the liquidity pool to ascertain an asset's price.
Furthermore, a majority of DeFi lending protocols mandate borrowers to provide
over-collateralization, instigating liquidation if a borrower's position
descends to under-collateralization. To maintain functional efficiency, lending
protocols integrate key parameters such as collateralization or liquidation
ratios.

\emph{Blockchain oracles.} 
Oracles provide real-world data to blockchains as they are tightly-closed systems and agnostic to such information.
Thus, oracles are critical for the smooth operation of DeFi protocols.
Specifically, price oracles furnish
indispensable information that has direct implications on both smart contract execution  and their results.
For instance, lending protocols use exact collateral asset prices to gauge
user risk profiles, and outdated or imprecise data may
precipitate financial losses.

In relation to oracle inputs, two distinct types of deviations can occur:
\emph{accuracy} and \emph{latency}. Accuracy deviations emerge when a value deviates from its
actual or true value, while latency deviations are identified when outdated
values are reported, a phenomenon that can, in turn, influence accuracy. These
deviations can originate from various sources such as intentional manipulation
of oracles to report distorted values, or unintentional data adjustments
embedded within smart contracts. Irrespective of their origins, such deviations
can result in incorrect operations within smart contracts.

\emph{Complexity of DeFi smart contracts.} 
\rev{Smart contracts implementing DeFi protocols such as lending, DEXes, and derivatives \revision{~\cite{aavev2,eulercontract, sol}} can be complex since they generally include loops to iterate through data structures representing various assets types or accounts managed by the protocols and calculate a sum, e.g., total assets or debts. \revision{This work highlights} that a typical \emph{Solidity} contract tends to include one loop per $250$ lines of cods and over $60\%$ of loops perform an accumulation~\cite{loopSummary}. }

%% file: overview.tex
\begin{figure}
\centering
\begin{lstlisting}
function borrowAllowed(address cToken, address bwr, uint brwAmt) external returns (uint) {
  ...
  uint surplus = hypotheticalLiquid(bwr,cToken,0,brwAmt);
  require(surplus > 0, "INSUFFICIENT_LIQUIDITY");
  ...
  return uint(Error.NO_ERROR);  }

function hypotheticalLiquid(address acct, CToken cToken, uint redTok, uint brwAmt) internal returns (uint) { 
  AccountLiquidityLocalVars memory v; 
  // Iterate over each asset in the acct
  CToken[] memory assets = accountAssets[acct];
  for (uint i = 0; i < assets.length; i++) {
    CToken asset = assets[i];
    (, v.cTokenBal, v.brwBal, v.exchRt) = 
    asset.getAccountSnapshot(acct);
    // Fetch asset price from oracle
    v.oraclePrice = oracle.getUnderlyingPrice(asset);
    v.collFact = markets[address(asset)].collFact; 
    v.tokensToDenom= v.collFact*v.exchRt*v.oraclePrice;
    v.sumColl = v.sumColl+ v.tokensToDenom* v.cTokenBal;
    v.sumBrwEfct= v.sumBrwEfct+ v.oraclePrice* v.brwBal;
    if (asset == cToken) {
      v.sumBrwEfct= v.sumBrwEfct+ v.tokensToDenom*redTok;      
      v.sumBrwEfct= v.sumBrwEfct+ v.oraclePrice*brwAmt;}}
  return v.sumColl - v.sumBrwEfct;                      }
\end{lstlisting}
\vspace{-2mm}
\caption{\emph{Compound} protocol borrow logic simplified.} \label{code:compound}
\vspace{-4mm}
\end{figure}

\begin{figure*}[!ht]
  \centering
\begin{lstlisting}[language=Summary, linewidth= \textwidth,mathescape=true]
$\sum_{a \in assets} ({collFact_a}*exchRt_a*p_a * cTokenBal_a) - (\sum_{a \in assets} (brwBal_a* p_a + c_a * (p_{brw} * brwAmt + collFact_r * exchRt_{r} *p_{r}))) > 0$
\end{lstlisting}
\vspace{-2mm}
\caption{Compound analysis summary}
\vspace{-3mm}
\label{fig:examplesummary}
\end{figure*}

\section{Example and Overview} \label{sec:example}
We present a motivating example of applying {\name} to analyze oracle
deviation in \emph{Compound}~\cite{compoundcontract}. Figure~\ref{code:compound} presents a simplified code
snippet from the \emph{Compound} smart contracts. \emph{Compound} is a decentralized borrowing
and lending protocol operating on the Ethereum blockchain. To borrow assets
from \emph{Compound}, a user deposits assets as collateral. The
total value of the collateral has to be significantly greater than the value
of the borrowed assets at any time. Whenever a user attempts to borrow assets,
\emph{Compound} calls \texttt{borrowAllowed} (line 1 in Figure~\ref{code:compound}) to enforce this policy. The function \texttt{borrowAllowed} in turn calls
\texttt{hypotheticalLiqu\-id} (line 8) to
calculate the difference (\emph{i.e.}, \texttt{surplus} at line 4) between the
adjusted value of the collateral assets (\emph{i.e.}, \texttt{v.sumColl} at line 20)
and the total value of the borrowed assets (\emph{i.e.}, \texttt{v.sumBrwEfct} at
line 21) for the given account \texttt{acct}. In the function, \emph{Compound} computes these two
values with the loop at lines 12-24. 
Each iteration of the loop handles one kind of asset in \emph{Compound} and updates
the two variables. Specifically, the loop computes \texttt{v.sumColl} as follows:

\begin{minipage}{0.9\columnwidth}
{\footnotesize
\begin{equation} \label{eqn:sumColl}
\begin{split} 
\sum_{a \in assets} \left( {collFact_a}*exchRt_a*p_a 
* cTokenBal_a \right)
\end{split}
\end{equation}}
\end{minipage}

\noindent where $\mathit{exchRt}_a$ is the exchange rate of the collateral asset $a$,
$p_a$ is the price of the asset $a$ fetched from an external oracle contract
(\texttt{v.oraclePrice} at line 20), $\mathit{cTokenBal}_a$ is the balance of the asset $a$, and
$\mathit{collFact}_a$ is a control variable smaller than one to
determine the enforced over-collateralization ratio for the asset.
The loop also computes \texttt{v.sumBrwEfct} as follows:

\begin{minipage}[l]{0.99\columnwidth}
{\footnotesize
\begin{equation}\label{eqn:RequiredBorrowEffect}
\hspace{-7mm}\sum_{a \in assets}(brwBal_a* p_a + c_a *(p_{brw} * brwAmt + collFact_r * exchRt_{r} *p_{r} 
  * redTok))
\end{equation}}
\end{minipage}

\noindent where $p_{brw}$ is the price of the asset $\mathit{brw}$ that the user wants to
borrow, $\mathit{r}$ is the asset the user wants to withdraw from its
collateral, $\mathit{brwBal}_a $ is the already borrowed balance of the
asset $a$, $\mathit{brwAmt}$ is the amount of the asset $\mathit{brw}$ a user
wants to borrow, and $\mathit{redTok}$ is the amount of asset
$r$ a user wants to redeem. $c_a = Int(a == cToken)$ is a binary representation 
of the condition at line 22, where $c_a = 1$ when the asset $a$ is $cToken$ and $c_a = 0$ otherwise. 
 Because \texttt{hypotheticalLiquid} can be invoked when a user
borrows assets or redeems collaterals, there are two different cases. The
second term corresponds to the borrowing case, while the last term corresponds to
the redeem case. 

\emph{Oracle values in Compound.}
The correctness of \emph{Compound} depends on the accuracy of the fetched
oracle price of each asset (line 17). Like many
other DeFi protocols, \emph{Compound} fetches oracle prices from multiple sources,
including centralized oracle service providers such as \emph{Chainlink}~\cite{chainlinkref} and
the trading price of the assets in decentralized protocols such
as \emph{Uniswap}~\cite{uniswapref}. However, values from these sources may deviate from 
ground truths. In fact, when a digital asset price is volatile,
obtaining the fair price of an asset is typically impossible. For instance, if
the prices in the equation \ref{eqn:sumColl} are inaccurately reported as
high, the value of users' collateral would increase, potentially leading the
protocol to execute borrowing transactions even when users are not sufficiently
collateralized. 

To tackle this issue, \emph{Compound} enforces additional margins for the positions of
each collateral asset and the margin sizes are determined by
\texttt{collFact}. \emph{Compound} empirically sets the collateral factor value lower
to enforce a larger margin on more volatile assets and sets the factor higher on
less volatile assets. Many other DeFi protocols have similar ad-hoc control
mechanisms to protect against oracle deviations. But there is a difficult
trade-off in how to set these control parameters appropriately. On one hand,
setting the parameters too relaxed would make the contracts vulnerable when
facing oracle deviations. On the other hand, setting the parameters too
restrictive would place additional collateral burden on users and make the
protocol unattractive.


\emph{Utilizing {\name}.} 
We now show how we apply {\name} to analyze \emph{Compound} to 
determine optimal control parameter values such as the collateral factor.
The user first identifies the interested operations in the source code. 
In our example, we identify \texttt{borrowAllowed} as entry point and 
\texttt{hypothetical\-Liquid}, which does
critical checks and computations when performing borrowing actions. 


\emph{Code analysis.} 
{\name} first analyzes the source code in Figure~\ref{code:compound} to
generate a symbolic expression for all variables in the constraint at line 4.
It starts with the entry function, replacing intermediate variables with their  
 computed expressions. For example, \texttt{surplus} is the returned value
of \texttt{hypothetical\-Liquid}. It is computed by subtracting \texttt{v.sumBrwEfct} from 
\texttt{v.sumColl}.
The expressions extracted by {\name} for \texttt{v.sumColl} and 
\texttt{v.sumBrwEfct} correspond to the mathematical formulas in Equations~\ref{eqn:sumColl} and \ref{eqn:RequiredBorrowEffect}, respectively.
\lstset{style=mystyle}

%
{\name} then generates the final symbolic expression for the safety
constraint at line 4 in Figure~\ref{fig:examplesummary}. Note that the terms in the final
expression are either loaded contract states (\emph{e.g.}, \texttt{v.brwBal}) or the return values
of external function calls (\emph{e.g.}, \texttt{getUnderlyingPrice}).

%
\emph{Loop summerization.}
{\name} handles the loop at lines 12-26 as follows.
With the observation that most loops in DeFi contracts perform fold operations, particularly accumulation,
{\name} summarizes the loop by identifying all accumulation performed and replacing the loop with
one or multiple compact expression(s).  
By replacing the variables and loops with compact expressions, the code summary
module returns a set of constraints to represent the smart contract's logic.
Constraints that are not affected by oracles will be ignored.
In this example, the constraint at line 4 in Figure~\ref{code:compound} will be extracted as the summary shown in Figure~\ref{fig:examplesummary}. Note that, in this summary, there are five vector
variables and three scalar variables. 

\emph{Formal model generation.}
The analysis results in Figure~\ref{fig:examplesummary} are then used to
construct a sound model of the safety constraint. Suppose we want to
investigate the behavior of the borrowing function in \emph{Compound} and identify
the price deviation limit when using the
default collateral factor (\emph{cf}) $0.7$, and a target collateral factor (\emph{cf}$^\prime$) 0.75.
Note that because of the deviation, the target value is always greater than the one 
configured in the contract. 
From the expression in Figure~\ref{fig:examplesummary},
we can derive the following simplified model:

\begin{minipage}{0.9\columnwidth}
{\footnotesize
\begin{equation*}\label{optprob}
\begin{aligned}
&\min_{\delta} \text{ }  \delta \\
&\text{s.t. } \forall \text{ } C, D, b, P, p, P_b, p_b > 0, 
 \frac{|P_i - p_i|}{P_i} < \delta, \frac{|P_b - p_b|}{P_b} < \delta,\\
&cf * \sum_{i}^{len} (C_i - D_i) * p_i - p_b * b > 0  \Rightarrow  cf' * \sum_{i}^{len} (C_i - D_i)*P_i - P_b * b > 0\\
\end{aligned}
\end{equation*}}
\end{minipage}

In this model, the variables 
 $\mathit{C}$, $\mathit{D}$, $\mathit{b}$ represent
 $\mathit{CollBal}$, $\mathit{brwBal}$ and 
 $\mathit{brwAmt}$ respectively. 
$\mathit{P}$ and $\mathit{P_b}$ stand for ground truth values while $\mathit{p}$ and $\mathit{p_b}$ stand for values reported by the oracle.
Note that because we are analyzing the borrowing case, the redeem amount is always zero and therefore the redeem related terms are simplified away.

\emph{Formal SMT solution.} Finally, we pass the model to the optimizer,
which iteratively calls an SMT solver to prove the constraint specified in the model above. 
Following, it returns the optimal $\delta$ if one is found. Then we can insert proper \texttt{require}
statements, for instance, restricting oracle deviation to be less than the value found, into the source code to ensure correct behavior. 


%% file: methodology.tex
\section{Design} \label{sec:design}
\rev{First, we introduce a simplified \emph{Solidity} language to help present our proposed analysis framework.
The language, shown in Listing~\ref{listing:simsol}, captures variable declarations, assignments, control-flow structures, and function calls.}

\rev{\emph{Contract, State, and Function.} A contract class has an identifier (\emph{id}) of String type (line~\ref{language:contract} in Listing~\ref{listing:simsol}). It encompasses a set of global states and functions. 
Each global state has a type and an identifier. We specify a function with its name, parameters and body which is constituted of a sequence of statements. 
The statements \emph{dec} and \emph{assign} (line~\ref{language:dec}) allow to declare a local variable and assign value to it, respectively.
The statement \emph{load} allows to read a contract's state.}

\rev{\emph{Control-Flow Structures.} 
Conditional branches are represented by \emph{IfThenElse} construct.} 
\rev{A \emph{for} loop is constituted of the loop iterator \emph{i}, upper-bound \emph{n} (for simplicity we omitted lower-bound), and a sequence of statements representing the loop body. 
A \emph{phi} instruction, denoted as \emph{$\phi_{id}$} is used to select values based on the control flow for a static single-assignment (SSA) smart contract. It can only appear at the beginning in a loop body or after an \emph{IfThenElse} construct.
The \emph{require} statement enforces smart contract constraints and its logic, and is crucial in our analysis.} 
\begin{lstlisting}[caption={\small \rev{Simplified Solidity Language.}}, captionpos=b, mathescape=true,label={listing:simsol},morekeywords={assign,load,dec,vcall,store, phi,state,func}]
Contract $\ssol{C}$ ::= contract($id$,$St^{*}$,$\ssol{F}^{*}$) $\label{language:contract}$
State $st$ ::= state($ty$,$id$)
Func $\ssol{F}$ ::= func($f$,$id^{*}$,$\ssol{S}^{*}$)
Type $ty$  ::= Int|Bool|Struct|Map|Array|Bytes|Address
Id   $id$  ::= String| $f$,$i$ $\in$ Id
Const $c$  ::= $n$ $\in$ Int|Bool|Bytes|Address
$Op$ ::= + | - | $\times$ | / | >= | > | < | <= | = 
Stmt $\ssol{S}$::= dec($ty$,$id$) | assign($id$,$\ssol{E}$) | $\label{language:dec}$
          load($id$,$st$) | require($\ssol{E}$) |
          phi($id_0$,$id_1$,$...$) |
          if($\ssol{E}_c$)then{$\ssol{S}_1^{*}$}else{$\ssol{S}_2^{*}$}$\Phi^{*}$ | 
          for($i$,$n$){$\ssol{S}^{*}$} |
          call($f$,$\ssol{E}^{*}$,$id^{*}$) | return($\ssol{E}^{*}$)
Expr $\ssol{E}$ ::= $c$ | $id$ | $id[\ssol{E}]$ | $id.id$ | $\neg \ssol{E}$ | $\ssol{E}_1\ Op\ \ssol{E}_2$
\end{lstlisting}

\rev{\emph{Function Calls.} 
Function calls are represented by the statement \emph{call($f$, $\ssol{E}^{*}$, $id^{*}$)}, where \emph{f} identifies the function, \emph{$\ssol{E}^{*}$} denotes the set of parameters, and \emph{id$^{*}$} represents the names of return values. 
Some function calls represent queries of states and to oracle. In our optimization problem, we consider those as free variables. 
Table \ref{tab:example} shows examples of statements from the \emph{Compound} protocol. }

\begin{table*}[h]
    \small
    \centering
    \caption{\small \rev{Example statements from Compound smart contracts in Solidity and simplified Solidity}}
    \vspace{-2mm}
    \label{tab:example}
    \begin{tabular}{l|c|c}
    \hline
    Solidity Code & require(surplus > 0, "INSUFFICIENT\_LIQUIDITY") & v.oraclePrice = oracle.getUnderlyingPrice(asset)\\\hline
    Simplified Solidity & require(surplus > 0) & call(oracle.getUnderlyingPrice, asset, v.oraclePrice)\\
\hline
\end{tabular}
\end{table*}

\subsection{Code Summary Overview}

\begin{figure}
    \centering
    \includegraphics[width=0.49\textwidth]{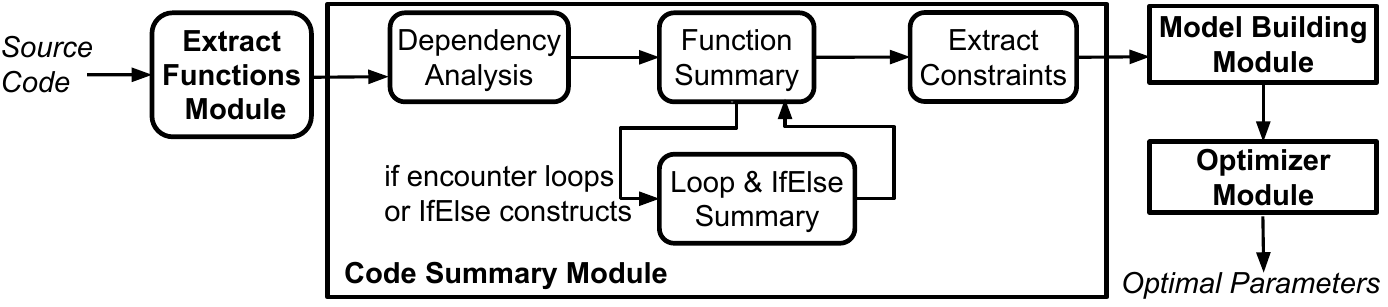}
    \vspace{-5mm}
    \caption{\rev{Overview of the proposed framework}}
    \label{fig:overview}
    \vspace{-2mm}
\end{figure}

Figure~\ref{fig:overview} presents an overview of the proposed framework.
The high-level procedure for the framework is shown in Algorithm~\ref{algo:main}.
The first step is to preprocess and simplify the source code to SSA form using the module \emph{ExtractFunc}. 
In \emph{ExtractFunc}, we identify the entry point of our analysis. 
Note that the entry point function must be a public function. 
In the case of \emph{Compound}, this entry-point is the \emph{borrowAllowed} function. 
Furthermore, the extracted functions are \emph{pure} functions, and they are invoked through the \emph{call} mechanism\footnote{In our implementation, we parse the abstract syntax tree (AST) of smart contracts and perform analysis on nodes in the extracted AST tree to identify pure functions. \revision{Pure functions are functions that have no side effects, \textit{i.e.}, do not modify contract's global states.}}. 
Then, \emph{CodeSummary} module extracts concise summaries including summaries of loops and conditionals and returns a list of constraints.
Next, \emph{BuildModel} module constructs an optimization model from the list of constraints for an SMT solver.
Lastly, \emph{SolveOpt} module solves the optimization model using the SMT solver.

\begin{algorithm}
\caption{\small\emph{Main} procedure. It takes in source code $sc$ of the benchmark.}
\label{algo:main}
\footnotesize
\begin{algorithmic}[1]
\Procedure{SummaryAnalysis}{$sc$}
    \State $ FuncObj \gets\Call {ExtractFunc}{sc}$     \label{algo:lineExtract}
    \State $ ConstLst \gets\Call{CodeSummary}{FuncObj}$ 
    \State $ M \gets\Call{BuildModel}{ConstLst}$ 
    \State $ OptVar \gets\Call{SolveOpt}{M}$
    \State $\textbf{return } OptVar$ 
\EndProcedure
\end{algorithmic}
\end{algorithm}
\vspace{-3mm}

\subsection{Code Summary Module}
\rev{The main tasks of the code summary module are: loop and oracle dependencies analysis,  extraction of symbolic expressions for variables, and constraints extraction. 
To compute a concise symbolic expression that can be passed to a solver, we summarize loops' bodies using the accumulation operator. 
Towards our goal, we introduce a domain-specific language (DSL) shown in Listing~\ref{listing:dsl}.}

\begin{lstlisting}[caption={\small \rev{Code summary DSL.}},captionpos=b,mathescape=true,  label={listing:dsl}]
aop     ::= +, -, *, /   ;   bop     ::= >, >=, <, <=, =
Id, i   ::= String       ;   lb, ub  ::= Int 
Const   ::= Int | Bool | Bytes | Address
Val $\dsl{V}$   ::= Id | Const | i | index($\dsl{V}_1$,$\dsl{V}_2$) | $\dsl{V}_1$($\dsl{V}_2$) |  
            $\dsl{V}_1$ aop $\dsl{V}_2$ | ret($\dsl{V}_2$, i)
Acc     ::= sum($\dsl{E}$, i, ub)           
Expr $\dsl{E}$  ::= Acc | $\dsl{V}$ | $\dsl{E}$ aop $\dsl{E}$ 
Constr $\dsl{C}$::= True | $\dsl{E}$ bop $\dsl{E}$ | $\neg \dsl{C}$
\end{lstlisting}
\vspace{-1mm}

\rev{The two main components of our DSL are \emph{$\dsl{E}$} for expressions and \emph{$\dsl{C}$} for Boolean constraint expressions. 
The \emph{$\dsl{E}$} type can either be an accumulation value (\emph{Acc}), a value (\emph{$\dsl{V}$}), or an arithmetic operation between two expressions. 
The \emph{$\dsl{C}$} type can either be the constant Boolean value \emph{True}, a comparison operation between two expressions, or a negation of another constraint.} 

\rev{\emph{Indexing and member-access.} We use the index operator \emph{index($\dsl{V}_1$, $\dsl{V}_2$)} to represent accessing an element from the array or map \emph{$\dsl{V}_1$} with the key \emph{$\dsl{V}_2$}.
The type of the \emph{$\dsl{V}_1$} must be either an array or map and the type of \emph{$\dsl{V}_2$} must be the same as the key's type of \emph{$\dsl{V}_1$}. This definition of the index operator allows nested indexing.
The member-access operator \emph{$\dsl{V}_1$($\dsl{V}_2$)} is used to represent accessing the field \emph{$\dsl{V}_1$} of the struct variable \emph{$\dsl{V}_2$}.}

\rev{\emph{Accumulation value.} 
To represent a loop's summary in our DSL, we use the accumulation operator \emph{sum($\dsl{E}$, i, ub)}, where \emph{i} is the iterator and \emph{ub} is the upper bound. 
The complexity of the summation is captured in the term \emph{$\dsl{E}$}, where it can be a complex mathematics formula involving multiple index and member-access operators.}

\rev{\emph{Return values of function calls.} We utilize \emph{ret($\dsl{V}$, i)} to indicate that \emph{$\dsl{V}$} is the return value of a \emph{pure} function which reads global states. 
When \emph{$\dsl{V}$} is a loop-dependent, \emph{i} represents the loop iterator, otherwise, \emph{i} is \emph{null}.
For example, at line 18 in Figure~\ref{code:compound}, \emph{v.oraclePrice} gets the value from the function \emph{oracle.getUnderlyingPrice} and the function is loop-dependent because of the argument \emph{asset}.
The generated summary is \emph{ret(oraclePrice(v), i)}.}

\subsubsection{Dependency Analysis} 
\rev{To determine expressions and require statements to include in our optimization model, we 
need to find which variables depend on the oracle price. 
Towards this, we propose a set of rules \emph{O1-O4} to infer this dependency and introduce the rule \emph{O5} to find guard statements that are oracle dependent. Moreover, to compute the loop summary we need to find which variables inside a loop body that depends on the loop iterator in order to account for them in the accumulation operator of our DSL. 
Thus, we also propose the set of rules \emph{L1-L4} to infer loop dependency. }

\rev{\emph{OD} denotes the set of expressions and statements that are oracle dependent. 
We do not distinguish between the two in \emph{OD}. \revision{The union operation for sets is denoted as $\uplus$.} \\
\noindent\textbf{O1:} If pure function reads from an oracle state then the identifier it reads to is oracle dependent.
If \emph{$\ssol{S}$ := call(f, $\ssol{E}$, id)} and \emph{Identifier(f) = oracle} where the helper function \emph{Identifier} checks whether the function is annotated as an oracle state getter, then \emph{id} $\in$ \emph{OD}.} 
\vspace{-1mm}
\rev{\begin{equation*} 
    \frac{\ssol{S} := call(f, \ssol{E}, id), Identifier(f) = oracle} { OD = OD \uplus  \{id,\ \ssol{S}\}}
\end{equation*}}
\rev{\noindent\textbf{O2:} If a statement reads from a global state that is oracle dependent and if a statement assigns an expression that is oracle dependent to a variable then the variable is oracle dependent.}
\vspace{-1mm}
\rev{\begin{equation*}
    \frac{\ssol{S} := load(id, st), st \in OD} { OD = OD \uplus  \{id, \ssol{S}\}} \quad 
    \frac{\ssol{S} := assign(id, \ssol{E}), \ssol{E} \in OD} { OD = OD \uplus  \{id,\ \ssol{S}\}}
\end{equation*}}
\rev{\noindent\textbf{O3:} For arithmetic and comparison expressions, if one of the operands is oracle dependent then the result is oracle dependent.}
\vspace{-1mm}
\rev{\begin{equation*}
    \frac{\ssol{E} = \ssol{E}_1\ op\ \ssol{E}_2, \ssol{E}_1 \in OD \lor \ssol{E}_2 \in OD} {  OD = OD \uplus  \{\ssol{E}\}}
\end{equation*}}
\rev{\noindent\textbf{O4:} For a function \emph{f}, if its parameter \emph{$\ssol{E}$} 
or one of its statements is oracle dependent, then the return value \emph{id} is oracle dependent.} 
\vspace{-1mm}
\rev{\begin{equation*}
    \frac{\ssol{S} := call(f, \ssol{E}, id), f := func(f,_,\ssol{S'}^{*}), \ssol{E} \in OD \lor \ssol{S'}^{*} \in OD} {  OD = OD \uplus  \{id,\ \ssol{S}\}}
\end{equation*}}
\rev{\noindent\textbf{O5:} A require statement is oracle dependent if its expression is.} 
\vspace{-1mm}
\rev{\begin{equation*}
    \frac{\ssol{S} := require(\ssol{E}), \ssol{E} \in OD} { OD = OD \uplus  \{\ssol{S}\}}
\end{equation*}}
\rev{We use \emph{L$_i$} to denote a \texttt{for} loop, with iterator \emph{i}, and corresponds to \emph{$\ssol{S}$ = for(i, n, $\ssol{S}^*$)}. \emph{LD$_i$} is the set of expressions that depend on \emph{L$_i$}.} 

\rev{\noindent\textbf{L1:} If an expression with an index that corresponds to the loop iterator or is loop dependent then the expression is loop dependent.}
\vspace{-2mm}
\rev{\begin{equation*}
    \frac{\ssol{E} = id[\ssol{E}], \ssol{E} = i \lor \ssol{E} \in LD_i } { LD_i = LD_i \uplus  \{\ssol{E}\}}
\end{equation*}}
\rev{\noindent\textbf{L2:} For arithmetic and comparison expressions, if one of the operands is loop dependent then the result is loop dependent.}
\vspace{-1mm} 
\rev{\begin{equation*}
    \frac{\ssol{E} = \ssol{E}_1 \ op \ \ssol{E}_2, \ssol{E}_1 \in LD_i \lor \ssol{E}_2 \in LD_i} {  LD_i = LD_i \uplus  \{\ssol{E}\}}
\end{equation*}}
\rev{\noindent\textbf{L3:} If a statement assigns an expression to a variable and this expression is loop dependent then the variable is loop dependent.}
\vspace{-1mm}
\rev{\begin{equation*}
    \frac{\ssol{S} := assign(id, \ssol{E}), \ssol{E} \in LD_i} { LD_i = LD_i \uplus  \{id\} }
\end{equation*}}
\rev{\noindent\textbf{L4:} If a statement invokes a function \emph{f} with a parameter \emph{$\ssol{E}$} that is loop dependent, then the return value \emph{id} is loop dependent.}
\rev{\vspace{-1mm}
\begin{equation*}
    \frac{\ssol{S} := call(f, \ssol{E}, id), \ssol{E} \in LD_i} { LD_i = LD_i \uplus  \{id\}}
\end{equation*}}

\subsubsection{Symbolic Value Extraction}
\rev{We now present a set of rules to generate a function summary, shown in Figure~\ref{fig:sumrules}. 
We use \emph{ExtractSummary} to refer to those rules. 
Specifically, \emph{ExtractSummary} takes a statement \emph{$\ssol{S}$} and an expression \emph{$\ssol{E}$}. 
It applies the effect of \emph{$\ssol{S}$} on \emph{$\ssol{E}$} and returns the updated expression \emph{$\ssol{E}'$}. We use \emph{$\summary{S}$} that maps a function \emph{f} to its summary.
Our approach uses a \emph{bottom-up} algorithm that begins from a \emph{return} or a \emph{require} statements. 
It adds the return expression \emph{$\ssol{E}$} of a function \emph{f} to the initially empty summary \emph{$\summary{S}(f)$}.}  
    
\rev{For an assignment, \emph{id = $\ssol{E}$}, we convert $\ssol{E}$ to its DSL using the procedure \emph{ConvDSL}. We substitute all occurrences of \emph{id} in \emph{$\summary{S}$(f) = $\dsl{E}$} with its computed DSL value, denoted as \emph{$\replace{\dsl{E}}{id}{ConvDSL(\ssol{E})}$}.}
 
\rev{For a function call, \emph{call(f', $\ssol{E}$, id)}, we generate a summary of \emph{f'($\ssol{E}$)}, denoted as \emph{S(f'($\ssol{E}$))}, and replace symbol \emph{id} with \emph{S(f'($\ssol{E}$))}.} 

\rev{The more involved summarization of loops and if-else statements are handled in the \emph{LpSm} and \emph{IfSm} procedures that we describe next.}

\begin{figure}
\centering
\small
\rev{
\begin{equation*} \label{r0} 
    \frac{pc(f):=\text{require $\ssol{E}$} \textbf{ or } pc(f):=\text{return $\ssol{E}$}, \summary{S}(f) = \perp, \dsl{E} := ConvDSL(\ssol{E})}
    {\summary{S} = \summary{S}[f \mapsto \dsl{E}] }
\end{equation*}
\begin{equation*}\label{r2}
    \frac{pc(f):= assign(id, \ssol{E}), \summary{S}(f) = \{\ssol{E}\}, \dsl{E}' := \replace{\dsl{E}}{id}{ConvDSL(\ssol{E})}}
    {\summary{S} = \summary{S}[f \mapsto \dsl{E}']}
\end{equation*}
\begin{equation*}\label{r3}
    \frac{pc(f):= call(f', \ssol{E}, id), \summary{S}(f) = \dsl{E}, \dsl{E}' := \replace{\dsl{E}}{id}{\summary{S}(f'(\ssol{E}))}}
    {\summary{S} = \summary{S}[f \mapsto \dsl{E}']}
\end{equation*}
\begin{equation*}
    \frac{pc(f):= for(i, n)\{\ssol{S}^*\},\dsl{E}' := LpSm(i, n, \ssol{S}^*,\summary{S}(f))}
    { \summary{S} = \summary{S}[f \mapsto \dsl{E}']}
\end{equation*}
\begin{equation*}
    \frac{pc(f):= if(\ssol{E}_c)then\{\ssol{S}_1^*\}else\{\ssol{S}_2^*\},\dsl{E}':= IfSm(\ssol{E}_c,\ssol{S}_1^*,\ssol{S}_2^*,\Phi^*,\summary{S}(f))}
    { \summary{S} = \summary{S}[f \mapsto \dsl{E}']}
\end{equation*}}
\vspace{-5mm}
\caption{\small \rev{Function summary extraction rules.}}\label{fig:sumrules}
\vspace{-3mm} 
\end{figure}

\rev{\emph{Loop Summary.} In Algorithm~\ref{algo:LoopSum}, we present the procedure \emph{LpSm}. 
\emph{LpSm} attempts to generate summaries for the symbols in the expression \emph{$\dsl{E}$}, in the form of accumulation or nested accumulation. For each symbol, we analyze the \emph{for} loop body in a bottom-up manner and perform symbolic substitution using \emph{ExtractSummary}.
We enumerate the symbols following their order of dependency (line~\ref{algo:dependency1}). For instance, in Listing~\ref{listing:lpsum}, since \emph{acc1} depends on \emph{acc} we enumerate \emph{acc1} before \emph{acc}. We use \emph{ps} to denote the "partial summary" of the symbol's value at an iteration \emph{i}.} 

\rev{We pattern match accumulation operations within \emph{ps} at line~\ref{EvalSum} and check whether a \emph{$\phi_{m}$} statement appears in the right-hand-side (rhs) precisely once. We also confirm that the rest of the expression, \emph{$\dsl{E}_s$}, is loop dependent using the loop-dependency set computed at line~\ref{AnalyzeDep}. 
If \emph{$\dsl{E}_s$} depends on \emph{m} which means that the loop violates the properties of an accumulation operation, we halt execution.}

\rev{For handling nested summations, we consider every computed symbol \emph{m1} with a summary \emph{v}.
We perform substitutions into the current summary \emph{ps} and adjust the inner-sum's upper bound at lines~\ref{FixIndex}-\ref{FixIndex2}. We also adjust previous summaries computed at line~\ref{Adjust}.}

\rev{If \emph{m} is an accumulator (\emph{findSum} is True), we remove assignments to \emph{m} from \emph{$\ssol{S}^*$} so that no substitution is performed for already computed symbols (lines~\ref{findSum1}-\ref{findSum2}). Finally, to compute the full summary at end of the loop, we set the upper bound of the outermost summation to match the loop's upper bound (line~\ref{upperboundn}).}

\begin{minipage}{\columnwidth}
\begin{lstlisting}[caption={\small \rev{Loop summary Example}},captionpos=b, mathescape=true,label={listing:lpsum}]
    acc: acc_0 + sum(index(A, j),j,b)
    acc1: acc1_0 + sum(acc_0 + sum(index(A, j),j,k),k,b)
    for (i =  0; i < b; i ++)  {
        acc' = phi(acc_0, acc$_{i-1}$)
        acc1' = phi(acc1_0, acc1$_{i-1}$)
        acc$_{i}$ =  acc' + A[i]
        acc1$_{i}$ = acc1' + acc$_{i}$        }
\end{lstlisting}
\end{minipage}
\vspace{-3mm}

\rev{In Listing~\ref{listing:lpsum}, we show an example of a nested sum and the computed complete summaries for both \emph{acc} and \emph{acc1}. }
\setlength{\textfloatsep}{8pt}
\begin{algorithm}
    \caption{\small\rev{\emph{LpSm} procedure. It takes loop parameters and an expression \emph{$\dsl{E}$} and returns an expression \emph{$\dsl{E}'$}. 
    \emph{M} stores symbols of \emph{$\dsl{E}$}, ordered based on \emph{$\ssol{S}^*$}.
    \emph{V} stores summaries of symbols in \emph{M}. 
    \emph{LD\_i} stores expressions that are loop-dependent. \emph{ApplyLDRules} updates \emph{LD\_i} using loop dependency rules. 
    }}
    \footnotesize
    \label{algo:LoopSum}
    \begin{algorithmic}[1]
    \Procedure{LpSm}{$i, n, \ssol{S}^*, \dsl{E}$}
        \State $V \leftarrow \{\},\ LD_i \leftarrow \{\}$
        \State $ \dsl{E}' \leftarrow \dsl{E}$
        \State \textbf{for each}\ \textbf{$m \in M$ in their order of dependency}  \label{algo:dependency1} 
        \State \ \ \ $p_s \leftarrow m$ 
        \State \ \ \ \textbf{for each}\ \textbf{$stmt \in reverse(\ssol{S}^*)$}                
        \State \ \ \ \ \ \  $p_s \leftarrow \Call{ExtractSummary}{stmt, p_s}$  
        \State \ \ \ \ \ \ $LD_i \leftarrow \Call{ApplyLDRules}{stmt, LD_i}$\ \textbf{using the rules L1-L4}  \label{AnalyzeDep} 
        \State \ \ \ \textbf{if}\ \textbf{$p_s \equiv "\phi_{m} + \dsl{E}_s" \land \dsl{E}_s \in LD_i $} \label{EvalSum}
        \State \ \ \ \ \ \ \textbf{if}\ \textbf{$(\dsl{E}_s \text{ depends on } m) $}
        \State \ \ \ \ \ \ \ \ \ $ \textbf{exit}$
        \State \ \ \ \ \ \ \textbf{else}
        \State \ \ \ \ \ \ \ \ \  $i' \leftarrow \Call{NewIndex}$ \label{NewIndex}
        \State \ \ \ \ \ \ \ \ \  $ps \leftarrow m_0 + sum(\replace{\dsl{E}_s}{i}{i'},i',i)$
        \State \ \ \ \ \ \ \ \ \ $findSum \leftarrow True$
        \State \ \ \ \textbf{for each}\ \textbf{$ (m1, v) \in V$} 
        \State \ \ \ \ \ \ \textbf{if}\ \textbf{$p_s \equiv ``m_0+sum(\dsl{E}_s,k,i)" \land m1 \in \dsl{E}_s \land$ $v \equiv ``v_0+sum(\dsl{E}_v,j,i)"$}
        \State \ \ \ \ \ \ \ \ \ $ps \leftarrow \replace{ps}{m1}{\replace{v}{i}{k}}$ \label{FixIndex}
        \State \ \ \ \ \ \ \textbf{if}\ \textbf{$p_s \equiv ``m_0+sum(\dsl{E}_s,k,i)" \land \phi_{m1} \in \dsl{E}_s \land$ $v \equiv ``v_0+sum(\dsl{E}_v,j,i)"$}
        \State \ \ \ \ \ \ \ \ \ $ps \leftarrow \replace{ps}{\phi_{m1}}{\replace{v}{i}{k-1}}$ \label{FixIndex2}
        \State \ \ \ \textbf{for each}\ \textbf{$ (m1, v) \in V$} 
        \State \ \ \ \ \ \ \textbf{if}\ \textbf{$p_s \equiv ``m_0 + sum(\dsl{E}_s,k,i)" \land \phi_{m} \in v$} \label{Adjust}
        \State \ \ \ \ \ \ \ \ \ $V[m1] \leftarrow \replace{v}{\phi_{m}}{\replace{ps}{i}{i-1}}$ 
        \State \ \ \ $ V[m] \leftarrow ps $
        \State \ \ \ \textbf{if}\ \textbf{$findSum$} \label{findSum1}
        \State \ \ \ \ \ \ $A \leftarrow A \setminus assign(m, \_) $ \label{findSum2}
        \State \textbf{for each}\ \textbf{$m \in M $}
        \State \ \ \ $ v \leftarrow \replace{V[m]}{i}{n}$  \label{upperboundn}
        \State \ \ \ $ \dsl{E}' \leftarrow \replace{\dsl{E}'}{m}{v}$ 
        \State $ \textbf{return } \dsl{E}'$
    \EndProcedure
    \end{algorithmic}
    \end{algorithm}

\rev{\emph{Handling conditional branches.}
To account for the different branches of an \emph{IfThenElse} construct, our summarization includes the condition in \emph{IfThenElse}. 
Algorithm~\ref{algo:IfElseSum} describes the procedure to summarize the effects of \emph{IfThenElse} construct. 
Similar to loops, we enumerate symbols in the order of dependency (line~\ref{algo:dependency1}). 
We then handle statements in reverse order and generate a summary for each symbol.
To combine summaries from both branches, we follow the \emph{phi} instruction and take the condition into account (line~\ref{combine}).}
\vspace{-1mm}
\begin{lstlisting}[caption={\small \rev{If-else branch example.}},captionpos=b,mathescape=true,label={listing:ifelse}]
    for(uint i = 0; i < b; i++)  {
        if (D[i]) { a1 = a1 + A[i] * B[i]; }
        else      { a2 = a2 + A[i] * C[i]; }    }
\end{lstlisting}
\rev{In the example shown above, we generate the following summaries.}
\begin{lstlisting}[language=Summary, label={listing:ifelsesum},mathescape=true]
a1=$\text{a1}_{0}$+sum(index(A,j)*index(B,j)*Int(index(D,j)),j,b)
a2=$\text{a2}_{0}$+sum(index(A,k)*index(C,k)*(1-Int(index(D,k))),k,b)
\end{lstlisting}
\rev{Expressions in the if branch are multiplied by the Boolean flag, \emph{D[i]}, while the ones in the else branch are multiplied by its complement. }
\subsubsection{Constraint Extraction} 
Our optimization model will consist of a set of constraints that are oracle-dependent or constraints over symbols that appear in or oracle-dependent constraints. 
The first set of constraints corresponds to guard (require) statements that are flagged as oracle-dependent 
using the oracle dependency analysis rules \emph{O1-O5}. 
The second set of constraints corresponds to guard statements that only contain symbols that appear in the set of oracle-dependent constraints.

\setlength{\textfloatsep}{8pt}
\begin{algorithm}
    \caption{\small\rev{\emph{IfSm} procedure. It takes \emph{IfThenElse} parameters and an expression \emph{$\dsl{E}$} and returns an expression \emph{$\dsl{E}'$}. \emph{$p_t$} and \emph{$p_e$} represent the partial summary for the two branches.
    \emph{$M_1$} and \emph{$M_2$} stores the symbols of \emph{$\dsl{E}$}. 
    Summaries of symbols in \emph{$M_1$} and \emph{$M_2$} are stored in \emph{V}.}}
    \label{algo:IfElseSum}
    \footnotesize
    \begin{algorithmic}[1]
    \Procedure{IfSm}{$\ssol{E}_c, \ssol{S}_{1}^{*}, \ssol{S}_{2}^{*}, {\Phi}^{*}, \dsl{E}$}  
        \State $ V \leftarrow \{\}$ 
        \State $ \dsl{E}' \leftarrow \dsl{E} $
        \State \textbf{for each}\ \textbf{$m_1 \in M_1, m_2 \in M_2$ in their order of dependency} \label{algo:dependency2}
        \State \ \ \ $p_{t} \leftarrow m_1,\ p_{e} \leftarrow m_2$
        \State \ \ \ \textbf{for each}\ \textbf{$stmt_{t} \in reverse(\ssol{S}_{1}^{*}), stmt_{e} \in reverse(\ssol{S}_{2}^{*})$}
        \State \ \ \ \ \ \  $p_{t} \leftarrow \Call{ExtractSummary}{stmt_{t}, p_{t}}  $ \label{updatetif} 
        \State \ \ \ \ \ \  $p_{e} \leftarrow \Call{ExtractSummary}{stmt_{e}, p_{e}} $ \label{updateelse}              
        \State \ \ \  $ V[m_1] \leftarrow p_{t}, V[m_2] \leftarrow p_{e}$ 
        \State \textbf{for each}\ \textbf{$\phi_{id} \equiv phi(id_1, id_2) \in {\Phi}^{*}$}
        \State \ \ \  $V[id] \leftarrow V[id_1] \times Int(\ssol{E}_c) + V[id_2] \times (1 - Int(\ssol{E}_c))$ \label{combine}
        \State \textbf{for each}\ \textbf{$m \in M$} 
        \State \ \ \  $ \dsl{E}' \leftarrow \replace{\dsl{E}'}{m}{V[m]}$            
        \State $ \textbf{return } \dsl{E}'$
    \EndProcedure
\end{algorithmic}
\end{algorithm}
\vspace{-3mm}
\begin{algorithm}
    \caption{\small \emph{BuildModel} Procedure. It takes in a list of constraints $ConstLst$ and returns a model $M$ for the SMT solver.}
    \label{algo:buildModel}
    \small
    \begin{algorithmic}[1]
    \Procedure{BuildModel}{$ConstLst$} 
        \State $Cv, Sd, Re, Ub \gets\Call{ExtractVars}{ConstLst} $ \label{buildmodel:1}
        \State $Cv, Sv, Re, Gt, \Delta \gets\Call{InitVar}{Cv, Sv, Re} $ \label{buildmodel:2}
        \State $C0, C1 \gets\Call{InitConst}{Cv, Sv, Re, Gt, \Delta}$ \label{buildmodel:3}
        \State $C \gets [C0, C1]$ \label{buildmodel:4}
        \State \textbf{for each}\ \textbf{$const \in ConstLst$} \label{buildmodel:5}
        \State \ \ \ $ [C_{Re}, C_{Gt}]\gets\Call{ConvertZ3}{const, Re, Gt, Cv, Sv, Ub}$  \label{buildmodel:6}
        \State \ \ \ $ \Call{append}{C, [C_{Re}, C_{Gt}] }$ \label{buildmodel:7}
        \State $\textbf{return }  M(\Delta, Cv, Sv, Re, Gt, C, Ub) $
    \EndProcedure
    \end{algorithmic}
    \end{algorithm}

\subsection{Model Generation}
We build a model \emph{M} from the extracted list of guard statements constraints. \emph{M} has $7$ parameters: Oracle prices values (\emph{Re}), ground truth prices values (\emph{Gt}), oracle deviation delta (\emph{$\Delta$}), the DeFi protocol's control variables (\emph{Cv}), \emph{e.g.}, margin ratio, state variables (\emph{Sv}), loops upper bounds (\emph{Ub}), and the set of constraints extracted \emph{C}. 

To generate \emph{M} from the guard statements, we first extract and initialize variables (lines~\ref{buildmodel:1}-\ref{buildmodel:2} in Algorithm \ref{algo:buildModel}). 
We also add two additional constraints \emph{C0}, and \emph{C1} to all models (lines~\ref{buildmodel:3}-\ref{buildmodel:4}). \emph{C0} states that \emph{Sv, Re, Gt} are greater than zero. \emph{C1} states that \emph{Re} deviates from \emph{Gt} by at most \emph{$\Delta$}. 
For each guard statements, we generate two constraints, one evaluated with ground truth values and the other one with oracle values (lines~\ref{buildmodel:5}-\ref{buildmodel:7}).

\subsection{Optimization}
Ideally, we are interested in finding some oracle deviations \emph{$\Delta$}, or control variables \emph{Cv}, such that the smart contracts "always behave correctly". In other words, for all inputs, given the deviated oracle price, the smart contracts should exhibit the same behavior as when given the ground truth price. For example, if we have a require statement: \emph{require(a > b)}, the corresponding constraint is \emph{a > b}, and assuming that one of the variables \emph{a, b} or both of them are functions of oracle inputs. We need to prove the following.

\vspace{1mm}
\begin{minipage}{0.9\columnwidth}
{\footnotesize
\begin{equation} \label{eqn:equivgt}
\hfill a(Re) > b(Re) \Rightarrow a(Gt) > b(Gt)
\end{equation}}
\vspace{-3mm}
{\footnotesize
\begin{equation} \label{eqn:equiv2}
a(Re) <= b(Re) \Rightarrow a(Gt) <= b(Gt)
\end{equation}}
\end{minipage}
\vspace{1mm}

Since we focus on the inputs when the transaction is not reverted, thus we only need to prove equation \ref{eqn:equivgt} (the require statement will revert the transaction if the lhs of equation \ref{eqn:equiv2} holds). 

\begin{algorithm}
    \caption{\small \emph{SolvOpt} Procedure. It takes in a model $M$, and returns the optimum parameters if found.}
    \label{algo:SolvOpt}
    \small
    \begin{algorithmic}[1]
    \Procedure{SolvOpt}{$M$} 
        \State $ConsList \gets\Call{simplifyConstraints}{M} $
        \State \textbf{while}\ \textbf{$\text{Stop condition not met}$}
        \State \ \ \ $res \gets\Call{Solv}{ConsList}$
        \State \ \ \ $OptVar \gets\Call{Update}{res}$
        \State $\textbf{return }  OptVar $
    \EndProcedure
    \end{algorithmic}
\end{algorithm}

There are several optimization problems that can be derived from the constraints. 
For example, we can solve for the maximum oracle deviation the protocol can tolerate given some control parameters \emph{Cv}, and \emph{Ub} (equation~\ref{optprob:delta}). 
That is, we maximize the oracle deviation delta such that for all inputs satisfying \emph{a > b}, given some predetermined control parameters. 
We can also give the model an expected delta and solve the optimization problem to find the optimum control parameters (equation \ref{optprob:cv}). In Algorithm \ref{algo:buildModel}, we give the procedure \emph{SolvOpt} that takes a model \emph{M} and iteratively queries a solver to find optimum parameters, or it reaches a timeout. 

\begin{minipage}{0.47\columnwidth}{\scriptsize
\begin{equation}\label{optprob:delta}
\begin{aligned}
&\max_{\Delta} \text{ } \Delta \\
\text{s.t. } 
& \forall \text{ }Sv > 0,
    \frac{|P_i - p_i|}{P_i} < \delta_i,\\
& a(Re, Sv, Cv) > b(Re, Sv, Cv) \Rightarrow \\
& a(Gt, Sv, Cv')>b(Gt, Sv, Cv')
\end{aligned}
\end{equation}}
\end{minipage}
\hfill
\vline
\hfill
\begin{minipage}{0.47\columnwidth}
{\scriptsize
\begin{equation}\label{optprob:cv}
\begin{aligned}
&\min_{Cv'}  \text{ }  Cv' \\
\text{s.t. } 
& \forall \text{ }Sv > 0,
    \frac{|P_i - p_i|}{P_i} < \delta_i,\\
& a(Re, Sv, Cv) > b(Re, Sv, Cv) \Rightarrow \\
&a(Gt, Sv, Cv') > b(Gt, Sv, Cv') 
\end{aligned}
\end{equation}}
\end{minipage}







%% file: experiment.tex
\vspace{2mm}
\section{Evaluation}\label{sec:results}
In this section, we evaluate the performance and effectiveness of {\name}, and present the evaluation results.
Specifically, we aim to answer the following research questions.
\begin{enumerate}[label=\textbf{RQ\,\arabic*:}, ref={RQ\,\arabic*},noitemsep,topsep=0pt]
    \item  \rev{Are current control parameters of Defi protocols safe under large oracle deviations?}
    \item  Can {\name} efficiently analyze various Defi protocols that use oracles?
    \item  Can {\name} assist developers to design safe Defi protocols that use oracles?
\end{enumerate}

\begin{table*}[ht]
\centering
\caption{\small Code summary module execution time.}
\vspace{-3mm}
\label{tab:codesum}
\small
\begin{tabular}{lccccccccc}
\hline
Protocol & \#requires & \#loops & \makecell[c]{compileTime (s)} & \makecell[c]{totalExecTime (s)} & \#vectorVars & \#otherVars & branch & dependency & oracle\\
\hline
\makecell[l]{Aave (borrow)} & 3 & 1 & 1.0558 & 1.0572 & 6 & 4 & \checkmark & \checkmark & Chainlink\\
\makecell[l]{Aave (liquidation)} & 1 & 1 & 0.6313 & 0.6350 & 5 & 1 & \checkmark & \checkmark & Chainlink\\
Compound  & 1 & 1 & 4.8403 & 4.8413 & 5 & 5 & \checkmark & \checkmark &OpenPriceFeed\\
Euler  & 1 & 1 & 2.4056 & 2.4063 & 5 & 2 & \checkmark & \checkmark &Uniswap\\
Solo  & 2 & 1 & 0.4704 & 0.4714 & 5 & 2 & \checkmark & \checkmark &Chainlink\\
Warp  & 1 & 2 & 1.5149 & 1.5156 & 4 & 2 & \checkmark & \checkmark &Uniswap\\
dForce  & \revision{1} & 2 & 1.3724 & 1.3746 & 6 & \revision{2} & \checkmark & \checkmark &Chainlink\\
Morpho & 1 & 2 & 8.5961 & 8.6002 & 7 & \revision{1} & \checkmark & \checkmark &Chainlink\\
TestAMM  & 1 & 0 & 0.1989 & 0.1992 & 0 & 4 & X & \checkmark &AMM-based\\
xToken & 0 & 0 & \revision{1.7244} & \revision{1.7247} & 0 & 4 & X & \checkmark &multiple source\\
Beefy  & 0 & 0 & 0.6730 & 0.6750 & 0 & 4 & X & \checkmark &depends on vault\\
\hline
\end{tabular}
\vspace{-3mm}
\end{table*}

\subsection{Implementation and Benchmarks}
We implement {\name} based on the \emph{Slither} static analysis tool~\cite{slitherref} with $1160$ lines of code in \emph{Python} for Solidity based smart contracts. To solve the optimization problems, we leverage the SMT solver \emph{Z3} \cite{z3SMT}.
\rev{Note that the constructs of the programming language in Listing~\ref{listing:simsol} that we used to present the main components of {\name} design are commonly found in other programming languages. Thus, {\name} implementation can also be extended to handle smart contracts written in other programming languages such as Vyper~\cite{vyper}.} 
\revision{The artifact is publicly available on Zenodo}~\cite{deng_2024_10436720}. 

\rev{We evaluate {\name} on $9$ DeFi protocols: \emph{Aave}, \emph{Compound}, \emph{Euler}, \emph{Solo}, \emph{Warp}, \emph{dForce}, \emph{Morpho}, \emph{Beefy}, and \emph{xToken}. 
Notably, this benchmark suite contains not only widely-used DeFi protocols according to DeFi industry database DefiLlama~\cite{defillamaref} but also that fell victim to oracle manipulation attacks.
To the best of our knowledge, \emph{Aave}, \emph{Compound}, \emph{Solo}, \emph{Morpho}, and \emph{Beefy} have not been victims to oracle manipulation attacks.
The protocols that were victims to oracle manipulation attacks are \emph{dForce}, \emph{Warp}, \emph{Euler}, and \emph{xToken}. 
We cover a wide range of protocols, including different types of lending protocols, yield aggregators, margin trading, and liquidity manager. 
We excluded several DeFi protocols, \emph{e.g.}, \emph{Inverse Finance}, \emph{CheeseBank}, \emph{JustLend}, \emph{Venus}, \emph{Benqi}, and \emph{Radiant}, that were forked from protocols in our benchmarks, \emph{e.g.}, \emph{Compound} and \emph{Aave}. }
We also evaluate {\name} on a fictional DeFi protocol \cite{calvwang9} developed to demonstrate oracle manipulation, and we call it \emph{TestAMM}.
All experiments are run on an AWS EC2 m5.2xlarge instance machine with 8vCPU, 32 GB memory, and 8TB SSD storage. 
\vspace{-2mm}
\subsection{Protocols' Response to Oracle Deviations}
\rev{To motivate {\name} and answer \textbf{RQ1}, we examine how oracle deviations impact the correctness
of DeFi protocols.} Specifically, we study historical oracle price deviations and the maximum
tolerance of each protocol with their default control parameter settings.

To narrow the scope of our study, we focus on the oracle price of ETH, the
native token of Ethereum network. \revision{We gather price updates for ETH/USD, USDT/ETH, USDC/ETH and DAI/ETH pair} on
\emph{Chainlink} and ETH/USDT pair on \emph{Uniswap}, where USDT, USDC and DAI are stable coins issued in
Ethereum which stay closely one to one with US dollar. \rev{We select \emph{Chainlink} and \emph{Uniswap} because they are very widely used oracles among DeFi protocols~\cite{defillamaoracle}, as also highlighted in the \emph{oracle} column of Table~\ref{tab:codesum}.} 

Because empirically oracle deviations often occur when a digital asset is highly volatile,
we study updates during the most volatile days of ETH for the two pairs
between 06/2020 and 09/2022. We compute the deviation 
as the difference between two consecutive updates on Uniswap. The rationale is that in normal
settings, the ground truth of an asset price is bounded by the values of two
consecutive updates. \revision{On Chainlink, we look for deviations within 33 minutes or 155 blocks window.}

Table \ref{tab:topdev} shows the top five deviations found. The first and the
third columns give the block number when the deviation is observed on \emph{Chainlink}
and \emph{Uniswap}, respectively. The second and fourth columns give the exact value
of deviations. 

Moreover, we study the maximum deviation allowed by each protocol.
Since the lending protocols require over-collateralization to cover borrowed or
leveraged positions, we define failure as when the borrowed value is more than
the collateral value of the user. For \emph{Aave}, \emph{Morpho}, \emph{Warp}, \emph{dForce}, \emph{Euler}, \emph{xToken}, and \emph{Beefy} 
we use the default control parameters of each protocol. 

Table \ref{tab:devi} shows the maximum tolerance of each protocol. 
Specifically, the first column gives the name of the protocol. The second
column specifies the parameter used in the experiment. The last column 
presents the maximum oracle deviation found. 

\begin{table}[h]\centering
    \caption{\small Top deviations observed on Chainlink and Uniswap.}\label{tab:topdev}
    \vspace{-3mm}
    \small
    \begin{tabularx}{\columnwidth}{lXXX}
    \hline
    Chainlink &Deviation &Uniswap &Deviation \\\hline
    \revision{11631223} &\revision{0.1390} &10314022 &0.4248 \\
    \revision{11631215} &\revision{0.1293} &10314022 &0.3351 \\
    \revision{11631215} &0.1260 &10326501 &0.2368 \\
    \revision{11631226} &0.1159 &10314022 &0.2356 \\
    \revision{11631248} &0.0994 &10326310 &0.1948 \\
    \hline
    \end{tabularx}
\end{table}

\begin{table}[h]
    \centering
    \caption{\small Deviation limit given specific control variables.}
    \label{tab:devi}
    \vspace{-3mm}
    \small
    \begin{tabularx}{\columnwidth}{lXX}
        \hline
        Protocol & CV & delta \\\hline
        Compound &cf = 0.7 &0.17 \\
        Aave &lth=0.85, ltv= 0.83 &0.08 \\
        \rev{Solo} &mp= 0.15, mr = 0.1 &0 \\
        Morpho &\revision{ltv} = 0.83 &0.09 \\
        Warp & cr = 2/3 &0.20 \\
        dForce & bf = 1, cf = 0.85 &0.08 \\
        Euler & bf = 0.91, cf = 0.9 &0.09 \\
        testAMM & cr = 0.7 &0.42 \\
        \rev{xToken} & fee = 0.02 & 0.02 \\
        \rev{Beefy} & fee = 0 & 0 \\
        \hline
    \end{tabularx}
\end{table}

\noindent \textbf{Answer to RQ1:}
We surprisingly found that the default control
parameters of the investigated protocols are \textit{not} enough
to protect these protocols against history oracle deviation. Specifically, for
protocols relying on \emph{Chainlink} price feed, \emph{e.g.}, \emph{Aave} and \emph{dForce}, with deviation
limits of 0.08, will suffer from under-collateralization given the greatest
deviation in Table~\ref{tab:devi}. \emph{Morpho} would encounter safety issues in
certain cases. \emph{Solo} is consistently at risk given the specific control parameters. 
\emph{Compound}'s Open Price Feed module relies on \emph{Chainlink} to update the price and
verify it by comparing it with \emph{Uniswap}'s average price. Thus, with a tolerance
of $0.17$, in some extreme cases, \emph{Compound} would execute incorrectly. \emph{Warp} and
\emph{Euler} use \emph{Uniswap} as price oracle and \emph{testAMM} relied on AMM-based oracles.
Deviations on \emph{Uniswap} are more significant, reaching a maximum value of $0.4248$.
While \emph{testAMM} would not suffer from under-collateralization in most cases given
the specific parameter, neither \emph{Warp} nor \emph{Euler} is safe given the oracle deviations. 

This finding means that the oracle deviation caused these protocols, at least
temporarily, to violate basic safety constraints such as over-collateralization. 
One consequence is, for example, that a malicious
attacker could send timely transactions during the deviation to borrow or
redeem assets with insufficient collaterals, extracting profits at the cost of
the protocol investors.

\rev{In the case of \emph{xToken} and \emph{Beefy}, where the protocol does not mandate over-collateralization,
 any price deviation leads to an immediate loss. The protocols charge fees for most operations proportional to the transaction amount. \emph{xToken} charges a maximum fee of $2\%$, while there is no deposit or withdraw fee in \emph{Beefy} vaults. Consequently, if we employ the fee as a control parameter, the maximum oracle deviation the protocol can tolerate will correspond to the percentage of the fee. Furthermore, in some cases, fees can be exploited in an attack. An example is the fee adjustment from 0.5\% to 0\%, contributing to the \emph{Yearn} attack in 2021~\cite{yearn,yearndisclosure}.}

\emph{Effect of Introducing Delay.} Introducing a delay is a widely
recommended approach to counteract oracle manipulation. An example of this
strategy can be found in \emph{MakerDao}'s OSM layer, which implements a one-hour
delay for price updates. \rev{This approach naturally introduces a deviation to the reported oracle price}. 
To evaluate this method, we conduct simulations using \emph{Chainlink} data 
and calculate the deviation from the current timestamp when a
one-hour delay is introduced. For instance, for the block with deviation of \revision{$0.1260$}, 
this strategy effectively reduces the deviation to \revision{$0.0779$}.

However, it is important to note that relying on a delayed price does not
guarantee a consistently smaller deviation. 
For instance, during the period from block $11541949$ to $11596096$, we notice an increase in the maximum deviation from $0.0329$ to $0.1525$ after applying the delay method. 
This finding underscores the complexity of defending against oracle manipulation and emphasizes the need to 
carefully consider the impact of delays in different contexts.

\vspace{-1mm}
\subsection{Effectiveness of {\name}}
To answer \textbf{RQ2} and assess the performance of {\name}, we run {\name} to
analyze the collected benchmarks. For the \emph{Aave} protocol, we apply {\name} to
the safety constraint in both borrowing and liquidation scenarios, which are
listed in rows one and two in Table~\ref{tab:codesum}, respectively. Notably, both Compound and Warp
protocols share the same set of constraints for their borrow and liquidation
operations. For \emph{Solo}, we apply {\name} on the safety constraint
that verifies whether a user's position is adequately collateralized. The
corresponding check is utilized in all operations, including liquidation,
within the protocol. For \emph{Euler}, we focus on the safety constraint responsible for checking liquidity in actions such as minting and withdrawal. As for \emph{dForce}, \emph{Morpho}, and \emph{TestAMM}, we analyze
the safety constraint of the borrow action. For \emph{xToken} and \emph{Beefy},
we focus on the constraint of the mint/deposit and burn/withdraw actions. 

Table~\ref{tab:codesum} presents the results of the experiment. The second and third columns present the 
number of "require" statements extracted and the occurrences of loops, respectively. 
We also include \emph{Slither} compilation time in the fourth column and
the total execution time in the fifth column. 
Moreover, we measure the number of vector variables and other variables (scalar)
in the constraints (columns five and six). We also present the features of each benchmark, including branching and dependent statements. 

\noindent \textbf{Answer to RQ2:}
Our results highlight the capability of {\name}. It successfully analyzed
all the protocols and their safety constraints in less than
$10$ seconds. We manually validated all the generated symbolic expressions.
For $8$ out of the $11$ cases, the contract code contains loops with dependencies
or branch conditions. For $7$ cases, the code contains more than one loop.
Although the code structures are difficult for standard analysis techniques,
our loop summary algorithm enables {\name} to handle all of them successfully.
Our loop summary algorithm is also fast, \emph{i.e.}, the majority of the execution
time is consumed by Slither to parse the code and generate AST.
\vspace{-1mm}
\subsection{Case Studies Analysis}

To answer \textbf{RQ3} and show how {\name} can help developers to design safe
protocols, we present case studies of applying {\name} on the nine benchmark protocols. For each case, we show how a user can use the symbolic
expressions obtained by {\name} to construct models to determine
appropriate values of control parameters when facing different


Timeout is set to be two minutes throughout the experiments. 

\emph{Compound} relies on the Open Price Feed module to access and retrieve price information critical to its operations. 
As discussed in Section \ref{sec:example}, the protocol implements a one-side risk control mechanism, \emph{i.e.}, uses a single control variable to govern its behavior.
Specifically, the control variable is known as the collateral factor. 
Normally, \emph{cf} is set to a value smaller than $1$, ensuring that the user's collateral value exceeds the borrowed value.
When \emph{cf} is greater than 1, the protocol allows under-collateralization, a situation generally considered undesired for lending protocols. We set the \emph{cf} to be 0.7 in the experiment, and consider three different oracle deviations. We run the search algorithm with a step size of $0.01$ for \emph{Ub=1} and $0.05$ for \emph{Ub=2}.  The results are shown in Table \ref{table:compoundliq}. 
The effective \emph{cf} achieved, \emph{i.e.}, \emph{cf}$^\prime$, is given in the second column.
The first column lists the parameter assignments,
the third column counts the number of free variables in the constraint and the optimization time is shown in the last column. 
We time out when we set \emph{bound=1, $\delta$=0.1}, and when we increase the bound \emph{Ub} to $3$.
We observe that the result would be the same if we use the same step size. 
Furthermore, it is reasonable to argue that the same \emph{cf}$^\prime$ would be optimal for \emph{Ub=3} as the search result should be independent of loop bounds.

Based on the results, if we expect an oracle deviation of $0.1$ and set \emph{cf = 0.7} (equivalent to $30$\% safety margin), the actual margin will be around $14$\%, \emph{i.e.}, \emph{cf}$^\prime$\emph{ = 0.86}. 
When there is no oracle deviation, we would achieve the exact margin specified in the protocol. 
This insight allows developers to understand how oracle deviations can impact the safety margin and offers guidance on parameter settings accordingly.
Furthermore, developers can add the corresponding constraint on oracle inputs which would guarantee the correctness. 

\begin{table}[h]
    \centering
    \caption{\small Compound with cf = 0.7 and ex = 1.}
    \label{table:compoundliq}
    \vspace{-3mm}
    \small
    \begin{tabularx}{\columnwidth}{lXXX}
        \hline
        Variables & cf$ ^\prime$ & \text{numVars} & time (s) \\\hline
       $\delta = 0.1, bound = 1$ &0.860 &5 &4.5486 \\
       $\delta = 0.01, bound = 1$ &0.720 &5 &0.1194 \\
       $\delta = 0.001, bound = 1$ &0.710 &5 &0.0794 \\
       \hline
        $\delta = 0.1, bound = 2 $& NA &9 &TO \\
        $\delta = 0.01, bound = 2$ &0.715 & 9 &0.3535 \\
        $\delta = 0.001, bound = 2$ &0.705 & 9 &0.2316\\\hline
    \end{tabularx}
\end{table}

\emph{Solo} project\cite{sol} is a marginal trading protocol of \emph{dXdY}, which uses \emph{Chainlink} as price oracle.
A desired property in \emph{Solo} protocol is that for all operations, accounts remain in a collateralized position. 
Besides, for liquidation operation, the protocol may not want to execute unnecessary liquidation, thus 
verifies that the account being liquidated is indeed under-collateralized before proceeding with the action.  
Protocol developers employ two control parameters to safeguard operations: the margin ratio (\emph{mr}) and the margin premium (\emph{mp}). 
For liquidation to safely happen, the following constraint (simplified), extracted by {\name}, must be met:
\vspace{-1mm}
\small
\begin{equation*}
    \frac{splyVal}{(1 - mp)} < brwVal * (1 + mp)  * (1 + mr)
\end{equation*}
\normalsize
where \emph{splyVal} represents the total collateral and \emph{brwVal} represents the total borrowed amount. 
These two variables are in the form of summation and are functions of oracle price input.

In the experiment, we set \emph{mr} to $0.1$ and \emph{mp} to $0.15$.
The results are shown in Table~\ref{table:sololiq}, similar to Table \ref{table:compoundliq}, 
except columns two and three presents the \emph{mp} and \emph{mr} achieved.  
When the bound \emph{Ub} is $1$, we achieve the margin set in the protocol.
However, as \emph{Ub} is increased to $2$, \emph{mr} achieved, denoted as \emph{mr}$^\prime$, also increases, resulting in a looser control effect. 
The experiment encountered a timeout when the \emph{Ub} was further increased to $3$.

\begin{table}[h]
    \centering
    \caption{\small Solo liquidation with mp = 0.15 and mr = 0.1.}
    \label{table:sololiq}
    \vspace{-3mm}
    \small
    \begin{tabularx}{\columnwidth}{lXXXX}
        \hline
        Variables & mp$^\prime$ & mr$^\prime$ & \text{numVars} & time (s) \\
        \hline
       $\delta = 0.1, bound = 1$ &0.15 & 0.10 &5 &0.0280 \\
       $\delta = 0.01, bound = 1$&0.15 & 0.10 &5 &0.0281 \\
       $\delta = 0.001, bound = 1$&0.15 & 0.10 &5 &0.0281 \\\hline
        $\delta = 0.1, bound = 2$&0.15 &0.35 &10 &1.1197\\
        $\delta = 0.01, bound = 2$ &0.15 &0.13 &10 &0.1454 \\
        $\delta = 0.001, bound = 2$ &0.15 &0.11 &10 &0.0888\\\hline
    \end{tabularx}
\end{table}

\emph{dForce} \cite{dforcecontract} is also a pool-based lending protocol and uses \emph{Chainlink} as price oracle. 
While \emph{dForce} also mandates over-collateralization, different from \emph{Compound}, \emph{dForce} designers enforce two-sided risk control, using $2$ control variables, the collateral factor (\emph{cf}) and the borrow factor (\emph{bf}). 
{\name} identifies the following safety constraint (simplified) in the smart contract to ensure collateralization:
\vspace{-1mm}
\small
\begin{equation}
    cf * \sum_{c=0}^{cl} (cb[c] * pc[c]) > \sum_{b=0}^{bl} \frac{(bb[b] * pb[b])}{bf}
\end{equation}
\normalsize
where \emph{cb} and \emph{bb} represent collateral and borrow balances, and \emph{pc} and \emph{pb} represent oracle prices. 
The constraint contains two summations, where bounds are represented by \emph{cl} and \emph{bl}, respectively.  

While for most assets, the protocol did not use \emph{bf} (\emph{bf=1}), 
we set \emph{cf}=$0.5$ and \emph{bf}=$0.7$ for the experiment purposes. 
As there are $2$ control variables to optimize, we fix one and search for the optimal value for the other. Table \ref{table:dforce} shows the experiment results. Columns \emph{cf}$^\prime$ and \emph{bf}$^\prime$ show 
the \emph{cf} and \emph{bf} achieved. 
We observe that the results obtained are independent of the bounds for all cases \emph{cl > 1, bl > 1}. 

\begin{table}[h]
    \centering
    \caption{\small dForce evaluation with cf = 0.5 and bf = 0.7.}
    \label{table:dforce}
    \vspace{-3mm}
    \small
    \begin{tabularx}{\columnwidth}{lXXXX}
        \hline
        Variables & cf$ ^\prime$ & bf$^\prime$ & \text{numVars} & time (s) \\\hline
       $\delta = 0.1, cl = 1, bl = 0$ & 0.5 & 0.7 &3 & 0.0264 \\
       $\delta = 0.01, cl = 1, bl = 0$ & 0.5 & 0.7 &3 & 0.0265 \\
       $\delta = 0.001, cl = 1, bl = 0$ & 0.5 & 0.7 &3 & 0.0263 \\
       \hline
        $\delta = 0.1, cl = 1, bl = 1$ & 0.5 & 0.856 & 6 & 3.8853\\
        $\delta = 0.01, cl = 1, bl = 1$ & 0.5 & 0.715 & 6 & 0.3996 \\
        $\delta = 0.001, cl = 1, bl = 1$ & 0.5 & 0.702 & 6 & 0.1091\\
        \hline
        $\delta = 0.1, cl = 1, bl = 1$ & 0.612  & 0.7 & 6 & 2.7787\\
        $\delta = 0.01, cl = 1, bl = 1$ & 0.511 & 0.7 & 6 & 0.3082\\
        $\delta = 0.001, cl = 1, bl = 1$ & 0.502 & 0.7 & 6 & 0.0892\\
        \hline
        $\delta = 0.1, cl = 2, bl = 1$ & 0.5 & 0.856 & 9 & 7.1360\\        
        $\delta = 0.01, cl = 2, bl = 1$ & 0.5 & 0.715 & 9 & 0.4700 \\
        $\delta = 0.001, cl = 2, bl = 1$ & 0.5 & 0.702 & 9 & 0.0875\\
        \hline
        $\delta = 0.1, cl = 2, bl = 1$ & 0.6115 & 0.7 & 9 & 3.8853\\
        $\delta = 0.01, cl = 2, bl = 1$ & 0.5110 & 0.7 & 9 & 0.3996 \\
        $\delta = 0.001, cl = 2, bl = 1$ & 0.5020 & 0.7 & 9 & 0.1091\\
        \hline
        $\delta = 0.1, cl = 2, bl = 2$ & 0.5 & 0.8560 & 12 & 9.0215\\        
        $\delta = 0.01, cl = 2, bl = 2$ & 0.5 & \revision{0.7145} & 12 & 0.3842 \\
        $\delta = 0.001, cl = 2, bl = 2$ & 0.5 & \revision{0.7015} & 12 & 0.1010\\
        \hline
        $\delta = 0.1, cl = 2, bl = 2$ & 0.6115 & 0.7 & 12 & 1.8996\\
        $\delta = 0.01, cl = 2, bl = 2$ & \revision{0.5105} & 0.7 & 12 & 0.6305 \\
        $\delta = 0.001, cl = 2, bl = 2$ & \revision{0.5015} & 0.7 & 12 & 0.1676\\
        \hline
    \end{tabularx}
\end{table}

\emph{xToken} \cite{xTokencontract} serves as a liquidity manager protocol, and it was the victim of an oracle manipulation attack.  
Specifically, the attacker was able to arbitrage because the protocol utilizes different price sources. 
The common attack vector involves three steps: first, minting or depositing the token; second, inflating the price of the minted token; and finally, withdrawing or burning the token. Other protocols such as yield aggregators are susceptible to such attacks.  
To mitigate these attacks, we propose an interface that compares the price at the time of withdrawal to the price at the time of minting. The equation we suggest for this comparison is as follows:
\small
\begin{equation}
\frac{| priceAtWithdraw - priceAtDeposit |}{priceAtDeposit} \leq \text{tol}
\end{equation}
\normalsize
Many existing protocols rely on a fixed tolerance ratio, which is ineffective when a big volume of tokens is traded. Thus, it is crucial to parameterize the variable \textit{tol} in order to take the amount of tokens traded into consideration. For example, we can parameterize \textit{tol} as $\frac{profitAllowance}{tokenVolume}$, which restricts the profit of a single transaction.  

We use {\name} to examine mint, burn, deposit, and withdraw, and
automatically extract the expression that approximates the price at withdraw and
deposit, \rev{shown in Table} \ref{table:yield}. The price at deposit is approximated as the value transferred to the
protocol and the token minted. Similarly, the withdraw price is represented by
the value transferred to the user divided by token burned. 
\begin{table}[h]
    \centering
    \caption{\small \revision{Expressions extracted for xToken.}}
    \vspace{-3mm}
    \label{table:yield}
    \begin{tabularx}{\columnwidth}{lXXXXX}
        \hline
        Protocol & mint & burn & numVars & time (s) \\\hline
        $xToken $& $\frac{etherContr}{mintAmt}$ & $\frac{valToSend}{tokToRedeem}$ &  4& 1.7247\\
        \hline
    \end{tabularx}
\end{table}

\emph{TestAMM} \cite{calvwang9} implements a simple lending protocol, which allows depositing USDC and borrowing ETH. It employs an AMM as a price oracle and uses a single risk control parameter named the collarerazation ratio (\emph{cr}). The simplified constraint for borrow action takes the form
\small
 \begin{equation} \label{eqn:equiv1}
  amount \leq ( deposits  * price ) * cr
\end{equation}
\normalsize

The result of applying OVer with \emph{cr} set to 0.7 is shown in Table \ref{table:ammresults}.
In this case, \emph{price} is computed using the constant product formula X * Y = K. 
Substituting \emph{price} with \emph{$X/Y$} yields identical outcomes.

\begin{table}[h]
    \centering
    \caption{\small testAMM evaluation with cr = 0.7.}
    \label{table:ammresults}
    \vspace{-3mm}
    \small
    \begin{tabularx}{\columnwidth}{lXXX}
        \hline
        Variables & cr$^\prime$ & numVars & time (s) \\\hline
        $\delta = 0.1$& 0.770  & 3 & 0.2403 \\
        $\delta = 0.01$& 0.710 & 3 & 0.0569 \\
        $\delta = 0.001$& 0.705 & 3 & 0.0423 \\
        \hline
    \end{tabularx}
\end{table}

\emph{Aave} \cite{aavev2} is one of the most popular pool-based lending protocols. 
Chainlink is employed as the primary price oracle, with a fallback oracle serveing as backup when queries to Chainlink fail. 
Aave enforces strict risk control mechanisms using two control variables, namely liquidation threshold \emph{lth} and loan to value ratio \emph{ltv}. 
For our borrow action experiment, we set \emph{lth}=$0.7$ and \emph{ltv}=$0.5$. The simplified constraints for the borrow action are shown in equation \ref{eqnaave} and \ref{eqnaave2}, 
\small
 \begin{equation} \label{eqnaave}
        \frac{\sum_{c\in assets}\left({pc[c] \times cb[c] \times lth}\right)}{\sum_{b\in assets}\left({pb[b] \times bb[b]}\right)} \geq 1
    \end{equation}

\begin{equation} \label{eqnaave2}
    \sum_{b \in assets}\left( {pb[b] \times bb[b]}\right) 
    + {pa \times amt } \leq \sum_{c\in assets} \left({pc[c] \times cb[c]} \times ltv \right)
\end{equation} 
\normalsize
where \emph{cb} and \emph{bb} represent collateral and borrow balances, and \emph{pc} and \emph{pb} represent oracle prices. 
\emph{pa} and \emph{amt} represent the price and the amount of the asset user wishes to borrow. 
 
The constraint for liquidation is the same as equation \ref{eqnaave} except the comparator changes to less than. 
We initially set the loop bound to one. When we increase the loop bound to two, the number of variables in the equations increase to 11, causing z3 to time out. 
The experiment for the liquidation call gives a similar result. A time-out occurs when we increase the loop bound to two.\\

\vspace{-3mm}
\begin{table}[h]
    \centering
    \caption{\small Aave borrow with lth = 0.7, ltv = 0.5 and bound = 1.}
    \label{table:aaveborrow}
    \vspace{-3mm}
    \small
    \begin{tabularx}{\columnwidth}{lXXXX}
        \hline
        Variables & lth$^\prime$ &  ltv$^\prime$ & numVars& time (s) \\\hline
        $\delta = 0.1$ &0.7 &0.612 &7 &22.2926 \\
        $\delta = 0.01$ &0.7 &0.511 &7 &2.5755\\
        $\delta = 0.001$ &0.7 &0.502 &7 &0.2063 \\
        \hline
    \end{tabularx}
\end{table}
\vspace{-5mm}

\begin{table}[h]
    \centering
    \caption{\small Aave liquidation with lth = 0.7 and bound = 1.}
    \label{table:aaveliq}
    \vspace{-3mm}
    \small
    \begin{tabularx}{\columnwidth}{lXXX}
        \hline
        Variables & lth$^\prime$ & numVars & time (s) \\\hline
        $\delta = 0.1$ &0.7 & 5 & 0.0331 \\
        $\delta = 0.01$ &0.7 & 5 &0.0321\\
        $\delta = 0.001$ &0.7& 5 &0.0359 \\
        \hline
    \end{tabularx}
\end{table}

\emph{Warp}~\cite{warpcontract} enables LP token-based borrowing and uses Uniswap V2 as a price oracle. 
The validation logic in the Warp protocol includes two loops, appearing in the constraint as well, with their respective bounds represented by \emph{cl} and \emph{bl}. 
The control variable, namely the collaterization ratio \emph{cr}, is hardcoded to be 2/3 within the protocol. The experiment result is shown in Table \ref{table:warp}. 
When we increase the bound \emph{cl} to two, the number of variables reaches nine, resulting in a timeout.

\begin{table}[h]
    \centering
    \caption{\small Warp evaluation with cr = 2/3 and ex = 1.}
    \label{table:warp}
    \vspace{-3mm}
    \small
    \begin{tabularx}{\columnwidth}{lXXXX}
        \hline
        Variables & cr$^\prime$ & \text{numVars} & time (s) \\\hline
    $\delta = 0.1, cl = 1, bl = 0$	& 0.815	& 6	& 3.6597\\
    $\delta = 0.01, cl =1, bl = 0$	& 0.681	& 6	& 0.3280\\
    $\delta = 0.001, cl=1, bl = 0$	& 0.670	&6	&0.0330\\
    \hline
    $\delta = 0.1, cl= 1, bl = 1$	&0.815	&9	&6.8087\\
    $\delta = 0.01,cl= 1, bl = 1$	&0.681	&9	&0.4758\\
    $\delta = 0.001, cl= 1, bl = 1$	&0.670	&9	&0.0467\\
    \hline
\end{tabularx}
\end{table}

\emph{Morpho} is a lending optimizer that implements a peer-to-peer layer over pool-based protocols such as Aave and Compound.
We study the Aave V3 based implementation \cite{morphoaave}. 
The simplified constraint is shown in equation \ref{eqn:morpho}, where \emph{cb} and \emph{bb} represent collateral and borrow balances, and \emph{pc} and \emph{pb} represent oracle prices. 
The constraint contains two summations, where bounds are represented by \emph{cl} and \emph{bl}, respectively.  
There is only one control variable in Morpho, \emph{ltv}, which is set to 0.7 in the experiment. 
In addition, we assume that the two bounds are equal. 
The result is shown in Table \ref{table:morpho}.
However, a timeout occurs when we increase the bound to two.

\small
\begin{equation}\label{eqn:morpho}
    \sum_{c=0}^{cl} (cb[c] * pb[c] * ltv) \geq \sum_{b=0}^{bl} (bb[b] * pc[b])
\end{equation}
\normalsize

\begin{table}[h]
    \centering
    \caption{\small Morpho evaluation with ltv = 0.7.}
    \label{table:morpho}
    \vspace{-3mm}
    \small
    \begin{tabularx}{\columnwidth}{lXXXX}
        \hline
        Variables & ltv$^\prime$ & \text{numVars} & time (s) \\\hline
        $\delta = 0.1, bound = 1$	&0.860	&6	&0.7198\\
        $\delta = 0.01, bound = 1$	&0.715	&6	&0.0947\\
        $\delta = 0.001,bound = 1$ &0.705	&6	&0.0478\\
        \hline
        $\delta = 0.1, bound = 2$ &	N/A	&12 & TO\\
        $\delta = 0.01, bound = 2$ &0.715 &12 &1.1773 \\
        $\delta = 0.001,bound = 2$ &0.705	&12	&0.1006\\
    \hline
\end{tabularx}
\end{table}

\emph{Euler} protocol \cite{eulercontract}, a pool-based lending protocol, shares the same control variables as the dForce protocol,
but \emph{Euler} only has one loop in its validation logic. 
The protocol enforces two-sided risk control.
We set the control variables, the collateral factor \emph{cf} to 1 and the borrow factor \emph{bf} to 0.7 in the experiment. 
By fixing \emph{cf}$ ^\prime$ at 1, we perform a search for the achieved value of \emph{bf}.
The experiment result aligns with that of the dForce protocol as expected.

\begin{table}[h]
    \centering
    \caption{\small Euler evaluation with bf = 0.7, cf = 1}
    \label{table:Euler}
    \vspace{-3mm}
    \small
    \begin{tabularx}{\columnwidth}{lXXXX}
        \hline
        Variables & bf$ ^\prime$ & \text{numVars} & time (s) \\\hline
        $\delta = 0.1, bound = 1$	&0.7	&4	&0.0312\\
        $\delta = 0.01, bound = 1$ &0.7&4	&0.0311\\
        $\delta = 0.001,bound = 1$ &0.7	&4	&0.0311\\
        \hline
        $\delta = 0.1, bound = 2$ &	0.860	&8 & 0.4572\\
        $\delta = 0.01, bound = 2$ &	0.715	&8 &0.4570 \\
        $\delta = 0.001,bound = 2$ &0.705	&8	&0.2674\\
    \hline
    \end{tabularx}
\end{table}
\vspace{-3mm}

\begin{table}[h]
    \centering
    \caption{\small Expressions extracted for Beefy}
    \label{table:yieldb}
    \vspace{-3mm}
    \begin{tabularx}{\columnwidth}{lXXXX}
        \hline
        Protocol & mint & burn & numVars & time (s) \\\hline
        $Beefy$	& $\frac{\_amount}{shares}$ & $\frac{r}{ shares}$ &  4&0.675\\
    \hline
    \end{tabularx}
\end{table}

\emph{Beefy} is a yield aggregator protocol, and we analyze its curve-kava-3pool \cite{beefyvault}. 
Similar to xToken, OVer helps analyze mint/deposit and burn/withdraw functions and approximate the price at withdraw and price at deposit.
The extracted expressions are shown in Table \ref{table:yieldb}.

\noindent \textbf{Answer to RQ3:}
Our study shows that the analysis results of {\name} can soundly capture the
logic of the target safety constraints for various kinds of DeFi protocols.
Given an oracle deviation ratio cap, a user can use the results of {\name} to
construct models to find optimal control parameters to guarantee
the desired safety property.

%% file: relatedwork.tex
\section{Related Work} \label{sec:related_work}
\emph{Automatic Analysis.}
A significant body of research has been dedicated to the automatic auditing of
smart contracts, utilizing classic methods such as fuzzing~\cite{jiang2018contractfuzzer, nguyen2020sfuzz, choi2021smartian, wang2020oracle, shou2023ityfuzz}, 
symbolic executions~\cite{mythrilref, mossberg2019manticore, luu2016making, liu2020towards, zheng2022park},
and static analysis~\cite{feist2019slither, tsankov2018securify, kalra2018zeus, tikhomirov2018smartcheck} to identify various
vulnerabilities. Researchers have also built verification tools that use formal
models to describe the intricate nature of these protocols and their
interactions~\cite{tolmach2021formal, bernardi2020wip, sun2020formal}. All the above
work focus on eliminating or nullifying implementation errors in smart
contracts. In contrast, we focus on the oracle deviation issue, which
is the input aspect of the contract. We propose the first sound analysis tool
to analyze oracle deviation in DeFi protocols.

Bartoletti, Massimo et. al. \cite{bartoletti2022formal} propose a simulation-based approach for
lending protocols, searching for optimal parameters to minimize non-repayable
loans. In contrast, {\name} works with existing \texttt{require} statements in the
code, eliminating the need for explicit safety property specifications.

\emph{Oracle Design and Runtime Mechanisms.} 
Extensive research has been conducted on DeFi protocols and the associated
attacks, with recent emphasis on flash loan attacks, as highlighted in the work
\cite{qin2021attacking, chen2022flashsyn}. Additionally, the manipulation of
oracles and price manipulation attacks have been extensively discussed. For
instance, this work \cite{9805499} demonstrates the vulnerability of lending
protocols that employ TWAP oracles to undercollateralized loan attacks.
\revision{Xue et. al.} \cite{xue2022preventing} suggest monitoring token changes in liquidity pools
to detect anomalous transactions and proposes using front-running as a defense
mechanism against such attacks. \revision{Wu et. al.} \cite{wu2021defiranger} propose a framework
for detecting oracle manipulation attacks through semantics recovery. An
algorithmic model is designed to estimate the safety level of DEX-based oracles
and calculate the cost associated with initiating price manipulation attacks
\cite{aspembitova2022oracles}.\revision{Wang et. al.} develop a tool that detects price manipulation 
vulnerabilities by mutating states ~\cite{wang2021promutator}. Several works
focus on the design of robust oracles and proving the properties of price
oracles \cite{dave_et_al:OASIcs.FMBC.2021.1}. While previous research has
primarily concentrated on the design of robust oracles and the detection of price
manipulation attacks, our work proposes promising analysis tools for smart
contracts to help developers to mitigate oracle deviation caused by such
attacks, operating under the assumption that oracles are unreliable.

\emph{Loop Summary.} 
The loop summary component of our work is closely related to a previous work
\cite{loopSummary} which proposes a DSL containing
map, zip, and fold operations and their variant to summarize \emph{Solidity} loops.
They use a type-directed search with an enumeration approach. However, after
multiple experiments we are not able to use their tool, \emph{Solis}, to implement the
loop summary component of our work since it does not handle loops that contain
\texttt{if-else} branches that require introducing Boolean flags in the summary.
Furthermore, during our experiments, we faced some loops without if branches
that require composition that cannot be handled using \emph{Solis}. For
example, the following loop requires composing the fold and zip operators on a
single statement which according to Section 4.3 in \cite{loopSummary} is not
supported in \emph{Solis}.
\vspace{-1mm}
\begin{lstlisting}[language=Solidity]
    for (uint i = 0; i < len ; i ++) {
        total += arr1[i] * arr2[i];   } 
\end{lstlisting}
\vspace{-1mm}
Also, as presented in Section 4.3 in \cite{loopSummary}, \emph{Solis} first generates a summary for a single statement and concatenates summaries through the sequence operator. Thus, it fails to handle dependent statements. 
\vspace{-3mm}
\begin{lstlisting}[language=Solidity]
    for (uint i = 0; i < len ; i ++)         {
        arr[i] * = 5; // S1
        total += arr[i]; //S2 depends on S1   }
\end{lstlisting} 
\vspace{-1mm}
In DeFi protocols, most loops perform fold operations and include complex
mathematical expressions.  Therefore, we develop a new loop summarization
algorithm that is tailored to DeFi smart contracts to address the above
issues.

%% file: threat.tex
\section{Threats to Validity}
\label{sec:threat}
One threat to the validity of our results is that {\name} might not be able to
analyze the source code of DeFi protocols beyond our benchmark set. To mitigate
this threat, we curated a diverse collection of benchmark protocols that manage
digital assets worth billions of dollars. Another potential threat is that we
focus on the five most volatile days in history to estimate the upper limit of
oracle deviations. However, even if we overlooked certain data points, it does
not undermine our surprising finding that the current control mechanisms in
many deployed DeFi protocols are inadequate for protecting the protocols
against oracle deviations.

%% file: conclusion.tex
\section{Conclusion} \label{sec:conclusion}

The integrity of decentralized finance protocols is frequently contingent on
the precision of crucial oracle values, such as the prices of digital assets.
In response to this, we introduced {\name}, the first sound
analysis tool that aids developers in constructing formal models directly from
contract source code. Our findings demonstrate that {\name} possesses the
capability to analyze a broad spectrum of prevalent DeFi protocols.
Intriguingly, with the assistance of {\name}, we discovered that many existing
DeFi protocols' control mechanisms, even with default parameters, fall short in
safeguarding the protocols against historical oracle deviations. This
revelation underscores the indispensable role of tools like {\name} and
advocates for more methodical strategies in the design of DeFi protocols.